\documentclass[aps,prl,twocolumn,nofootinbib,longbibliography,amsfonts,amssymb,amsmath,superscriptaddress]{revtex4-2}
\usepackage[T1]{fontenc}
\usepackage[dvipsnames, table]{xcolor}
\usepackage{IEEEtrantools}
\usepackage{mathrsfs}
\usepackage{amsthm}
\usepackage{enumitem}
\usepackage{varwidth}
\usepackage[normalem]{ulem}
\usepackage{graphicx}
\usepackage{stackengine}
\usepackage{cprotect}
\usepackage[caption=false]{subfig}
\usepackage{multirow}
\usepackage[USenglish]{babel}
\usepackage{microtype}
\usepackage{xspace}
\usepackage{placeins}

\usepackage{siunitx}
\usepackage{tabularx}
\usepackage{booktabs}
\usepackage{array}
\usepackage{csvsimple}
\usepackage{xstring}
\usepackage[pdftex,colorlinks=true,allcolors=blue1]{hyperref}
\definecolor{blue1}{HTML}{1065AB} 
\definecolor{red2}{HTML}{B31529} 

\usepackage{etoolbox}
\usepackage{pgf} 
\definecolor{maxperc}{HTML}{1065AB} 
\definecolor{blue2}{HTML}{3A93C3} 
\definecolor{five_perc}{HTML}{8EC4DE} 
\definecolor{blue4}{HTML}{D1E5F0} 
\definecolor{one_perc}{HTML}{F9F9F9} 
\definecolor{peach1}{HTML}{FEDBC7} 
\definecolor{towards_one_perc}{HTML}{F6A482}  
\definecolor{red1}{HTML}{D75F4C} 
\definecolor{zero_perc}{HTML}{B31529} 

\newcommand*{\opacity}{40}

\newcommand*{\minval}{0.0}     
\newcommand*{\valA}{0.8}       
\newcommand*{\valB}{1.0}      
\newcommand*{\valC}{5.0}      
\newcommand*{\maxval}{70.0}    
\newcommand{\gradient}[1]{%
	\pgfmathsetmacro{\value}{#1}%
	\pgfmathsetmacro{\value}{min(max(\value,\minval),\maxval)}%
	
	\ifdim\value pt<\valA pt
	\pgfmathsetmacro{\pctraw}{100*(\value - \minval)/(\valA - \minval)}%
	\pgfmathparse{int(round(\pctraw))}%
	\edef\colorcmd{\noexpand\cellcolor{towards_one_perc!\pgfmathresult!zero_perc!\opacity}}%
	
	\else\ifdim\value pt<\valB pt
	\pgfmathsetmacro{\pctraw}{100*(\value - \valA)/(\valB - \valA)}%
	\pgfmathparse{int(round(\pctraw))}%
	\edef\colorcmd{\noexpand\cellcolor{one_perc!\pgfmathresult!towards_one_perc!\opacity}}%
	
	\else\ifdim\value pt<\valC pt
	\pgfmathsetmacro{\pctraw}{100*(\value - \valB)/(\valC - \valB)}%
	\pgfmathparse{int(round(\pctraw))}%
	\edef\colorcmd{\noexpand\cellcolor{five_perc!\pgfmathresult!one_perc!\opacity}}%
	
	\else
	\pgfmathsetmacro{\pctraw}{100*(\value - \valC)/(\maxval - \valC)}%
	\pgfmathparse{int(round(\pctraw))}%
	\edef\colorcmd{\noexpand\cellcolor{maxperc!\pgfmathresult!five_perc!\opacity}}%
	\fi\fi\fi
	
	\colorcmd\rule{0pt}{2.5ex}\hspace{4pt}#1~\%\hspace{3pt}
}

\newcommand{\dingo}{\textsc{Dingo}\xspace}
\newcommand{\dingoT}{\mbox{\textsc{Dingo-T1}}\xspace}
\renewcommand{\sec}[1]{\emph{#1.---}}
\DeclareSIUnit \parsec {pc}

\begin{document}
	
	\title{Flexible Gravitational-Wave Parameter Estimation with Transformers}
	
	\author{Annalena Kofler}
	\email{annalena.kofler@tuebingen.mpg.de}
	\affiliation{Max Planck Institute for Intelligent Systems, Max-Planck-Ring 4, 72076 T\"ubingen, Germany}
	\affiliation{Max Planck Institute for Gravitational Physics (Albert Einstein Institute), Am M\"uhlenberg 1, 14476 Potsdam, Germany}
	
	\author{Maximilian Dax}
	\affiliation{Max Planck Institute for Intelligent Systems, Max-Planck-Ring 4, 72076 T\"ubingen, Germany}
	\affiliation{ELLIS Institute T\"ubingen, Maria-von-Linden-Straße 2, 72076 T\"ubingen, Germany}
	\affiliation{T\"ubingen AI Center, Maria-von-Linden-Straße 1, 72076 T\"ubingen, Germany}
	
	\author{Stephen R. Green}
	\affiliation{Nottingham Centre of Gravity \& School of Mathematical Sciences, University of Nottingham, University Park, Nottingham, NG7 2RD, United Kingdom}
	
	\author{Jonas Wildberger}
	\affiliation{ELLIS Institute T\"ubingen, Maria-von-Linden-Straße 2, 72076 T\"ubingen, Germany}
	
	\author{Nihar Gupte}
	\affiliation{Max Planck Institute for Gravitational Physics (Albert Einstein Institute), Am M\"uhlenberg 1, 14476 Potsdam, Germany}
	\affiliation{Department of Physics, University of Maryland, College Park, MD 20742, USA}
	
	\author{Jakob H. Macke}
	\affiliation{Max Planck Institute for Intelligent Systems, Max-Planck-Ring 4, 72076 T\"ubingen, Germany}
	\affiliation{Machine Learning in Science, University of T\"ubingen \& T\"ubingen AI Center, 72076 T\"ubingen, Germany}
	
	\author{Jonathan Gair}
	\affiliation{Max Planck Institute for Gravitational Physics (Albert Einstein Institute), Am M\"uhlenberg 1, 14476 Potsdam, Germany}
	
	\author{Alessandra Buonanno}
	\affiliation{Max Planck Institute for Gravitational Physics (Albert Einstein Institute), Am M\"uhlenberg 1, 14476 Potsdam, Germany}
	\affiliation{Department of Physics, University of Maryland, College Park, MD 20742, USA}
	
	\author{Bernhard Schölkopf}
	\affiliation{Max Planck Institute for Intelligent Systems, Max-Planck-Ring 4, 72076 T\"ubingen, Germany}
	\affiliation{ELLIS Institute T\"ubingen, Maria-von-Linden-Straße 2, 72076 T\"ubingen, Germany}
	
	\begin{abstract}
		Gravitational-wave data analysis relies on accurate and efficient methods to extract physical information from noisy detector signals, yet the increasing rate and complexity of observations represent a growing challenge. 
		Deep learning provides a powerful alternative to traditional inference, but existing neural models typically lack the flexibility to handle variations in data analysis settings.
		Such variations accommodate imperfect observations or are required for specialized tests, and could include changes in detector configurations, overall frequency ranges, or localized cuts. 
		We introduce a flexible transformer-based architecture paired with a training strategy that enables adaptation to diverse analysis settings at inference time. 
		Applied to parameter estimation, we demonstrate that a \emph{single} flexible model---called \dingoT---can (i)~analyze 48 binary black holes from the third LIGO-Virgo-KAGRA Observing Run under a wide range of analysis configurations, (ii)~enable systematic studies of how detector and frequency configurations impact inferred posteriors, and (iii)~perform inspiral-merger-ringdown consistency tests probing general relativity. 
		\dingoT also improves median sample efficiency on real events from a baseline of 1.4\% to 4.2\%. Our approach thus demonstrates flexible and scalable inference with a principled framework for handling missing or incomplete data---key capabilities for current and next-generation observatories.
	\end{abstract}
	
	\maketitle

	\sec{Introduction}Gravitational waves~(GWs) provide unique access to the strong-field dynamics of compact objects and have transformed our ability to probe astrophysics, cosmology, and fundamental physics~\cite{GBM:2017lvd, lvk_gw170817:2018, lvk_gwtc3_populations:2023, lvk_gwtc3_tests_gr:2025, lvk_gwtc3_expansion:2023, lvk_gwtc4_populations:2025,  lvk_spec_tests_gr_gw250114:2025, lvk_gwtc4_expansion:2025}. 
	Drawing scientific conclusions from GW~observations requires accurate and efficient data analysis, yet conventional Bayesian techniques~\cite{bilby,ashton_bilby:2021} are computationally expensive and difficult to scale to the growing volume and complexity of data. 
	Machine learning offers a promising solution by leveraging neural networks to process strain data and predict physical quantities, and has seen applications to detection~\cite{gebhard:2019, Marx:2024wjt, Koloniari:2024kww}, noise classification~\cite{Razzano:2018fxb}, source localization~\cite{Chatterjee:2022ggk, Chatterjee:2022dik}, and parameter estimation~(PE)~\cite{Gabbard:2019rde, Chua:2019wwt, green_anf:2020, green_full_npe:2020, Delaunoy:2020zcu, dax_dingo:2021, dax_is:2022, Bhardwaj:2023xph, Chatterjee:2024pbj, dax_bns:2025, vanStraalen:2024xiq}. 
	For PE, neural simulation-based inference~(SBI, \cite{cranmer_sbi:2020}) has reduced analysis times for compact binary coalescences to the order of a few seconds---crucial for handling increased event rates~\cite{ligo_gwtc-21:2024,ligo_gwtc-3:2023} and enabling rapid multi-messenger follow-up~\cite{GBM:2017lvd}.
	
	This speed-up arises from front-loading computational costs: once trained on millions of simulated data sets, neural networks can analyze new observations at negligible expense. 
	However, conventional neural architectures operate on a fixed input dimension, requiring the data representation to be fixed prior to training. 
	A trained model is therefore tied to specific detector combinations and data-conditioning settings---a major barrier to widespread adoption. 
	Existing models cannot readily adapt to detectors being offline, to restricted frequency bands used to mitigate noise contamination, or to data gaps expected in LISA observations~\cite{burke:2025}.
	
	In this \emph{Letter}, we propose a neural architecture and training strategy that enable flexible analysis of heterogeneous GW~data. 
	The key idea is to process strain data using a transformer encoder and exploit its ability to handle input sequences of arbitrary length~\citep{vaswani:2017}. 
	We partition data sets into sequences of short strain segments called tokens, and randomly drop tokens during training to mimic frequency cuts and missing detectors. 
	Through the self-attention mechanism, the model learns to identify correlations between strain segments, capturing global dependencies across detectors and frequencies, and to marginalize over missing data. 
	Although transformers have been used for a range of GW~tasks~\citep{jiang:2022, shi:2024, wang_waveformer:2024, chatterjee:2024, khalouei:2025, joshi:2025, papalini:2025}, previous work has not leveraged their inherent flexibility in handling variable-length or incomplete data. 
	Our approach is inspired by works on missing data in SBI~\citep{shukla:2021, wang_missing_data:2024, gloeckler:2024, verma:2025}.
	
	\begin{figure*}[t]
		\centering
		\includegraphics[width=\textwidth]{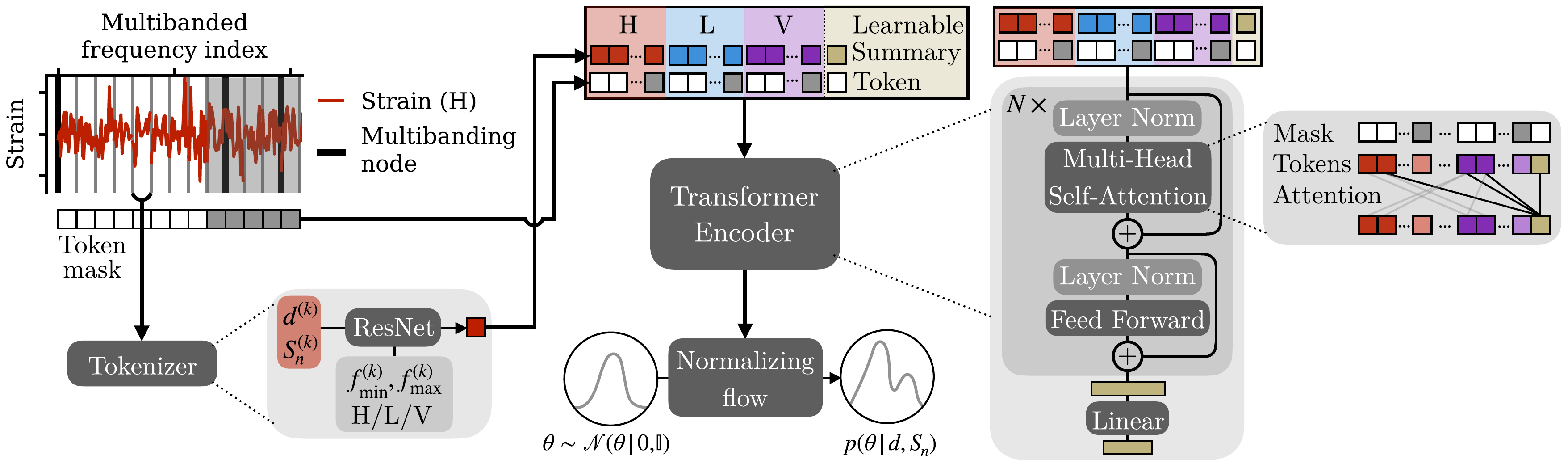}
		\caption{Overview of \dingoT. After multibanding, data and detector PSD segments~$(d^{(k)}, S_n^{(k)})$ are mapped by a shared tokenizer into tokens that also encode the frequency range~$(f^{(k)}, f^{(k+1)})$ and detector identity~$I$. The sequence of tokens is masked, augmented with a learnable summary token, and processed by the transformer encoder. Through self-attention, the summary token aggregates information from all unmasked segments. The final summary token is projected via a linear layer to a 128-dimensional feature vector, which conditions a normalizing flow to model the posterior $p(\theta|d, S_\mathrm{n})$ over source parameters.}
		\label{fig:architecture}
	\end{figure*}
	
	We integrate the transformer encoder into the \dingo framework for GW~PE, yielding a flexible model that we call \dingoT~(\dingo~Transformer, version~1; see Fig.~\ref{fig:architecture}). 
	We demonstrate its versatility and performance through injection studies and analyses of real data. 
	In particular, with a single network, we reanalyze 48~out of 84~binary black holes~(BBHs) from the LIGO-Virgo-KAGRA (LVK)~\cite{ligo:2014,virgo:2014,kagra:2020} third observing run~(O3)~\citep{ligo_gwtc-21:2024, ligo_gwtc-3:2023, ligo_data_release_gwtc21_zenodo:2022, ligo_data_release_gwtc3_zenodo:2021} that are consistent with our prior (generic spins, component masses $m_1, m_s \in [10, 120]~M_\odot$, and luminosity distance $d_L \in [0.1, 6]~\si{\giga \parsec}$). 
	Together, these events span 17~combinations of detectors and frequency ranges. 
	Compared to baseline neural posterior estimation~(NPE) networks with residual network encoders~\citep{dax_dingo:2021}, \dingoT improves median importance sampling efficiency from 1.4\% to 4.2\%. 
	We further analyze all events using all possible LIGO-Virgo detector combinations. 
	Such studies are extremely costly for standard methods, but can be used to identify data systematics. 
	As a final demonstration, we use frequency cuts to perform inspiral-merger-ringdown consistency tests on seven events where \dingoT yields reliable posterior estimates. 
	Previously, conducting all of these experiments would have required training 94~different models to handle the various detector combinations, and frequency ranges for events as well as tests of general relativity.
	
	Our study thus represents a step towards generalized inference models in GW~astronomy, enabling flexible choice of detector and frequency settings without retraining. 
	This flexibility opens new prospects for analyzing large catalogs~\cite{ligo_gwtc-21:2024,ligo_gwtc-3:2023}, facilitating rapid assessments of systematics and consistency across diverse data-analysis settings~\cite{payne:2022, udall:2025}, as well as more robust approaches to missing data.
	In addition, our transformer model \dingoT\footnote{Weights for \dingoT and \dingo baseline~\citep{kofler_dingoT_zenodo}, as well as corresponding code are publicly available: \href{https://github.com/dingo-gw/dingo-T1}{https://github.com/dingo-gw/dingo-T1}. The model was trained taking~\citep{talbot_window_fix:2025} into account.} serves as a powerful and general encoder for GW~data which could be fine-tuned on other tasks such as GW~detection, providing a foundation for efficient and scalable GW~science.
	
	\sec{Architecture}For NPE~\cite{papamakarios:2016, lueckmann:2017, greenberg_lfi:2019}, one trains a neural network $q(\theta|d)$ to estimate the Bayesian posterior~$p(\theta|d)$ over parameters~$\theta$ given data~$d$. 
	In the context of GW single-event inference, $\theta$ denotes the source parameters, e.g., of a black hole binary, and $d$ denotes the strain data measured at the detectors~\cite{LIGOScientific:2019hgc}. 
	For \dingo, we also augment the strain data with an estimate of the noise power spectral density~(PSD), which varies from event to event~\cite{dax_dingo:2021}. 
	Data are generally of high dimension \mbox{($\gg 10^4$)}, so $d$~is first mapped by an embedding network~$f_{\phi}$ to a lower-dimensional representation. This compressed summary is then passed to a density estimator~$\hat{q}_{\varphi}$ to model the posterior, $q(\theta|d)=\hat{q}_{\varphi}(\theta|f_{\phi}(d))$.
	The learnable parameters $\{\phi,\varphi\}$ are optimized jointly during training.
	
	Our core architectural change is to replace the traditional embedding network with a transformer encoder (Fig.~\ref{fig:architecture}), which naturally handles sequences of variable length. We divide the strain data sets into small segments, which a tokenizer network maps to token embeddings (Fig.~\ref{fig:architecture} left). These embeddings form a sequence that the transformer encoder processes. 
	We now describe the tokenization scheme, model architecture, and training strategy; full technical details appear in the Supplemental Material~\citep{supplemental}.
	
	We represent GW data $d = \{d_I\}_{I=\mathrm{H,L,V}}$ as frequency series from the Hanford~(H), Livingston~(L), and Virgo~(V) detectors covering $[20, 1810]$~\si{\hertz}. 
	We use a multi-banded (non-uniform) frequency grid, with coarser resolution at higher frequencies, designed to capture the chirping morphology of GW signals, and achieve an initial compression of the data~\citep{vinciguerra_mb:2017,morisaki:2021,dax_bns:2025}. To tokenize, we partition the grid into $K$ equal-length segments with boundary nodes $(f^{(k)})_{k=0}^{K+1}$, such that each interval $\big(f^{(k)}, f^{(k+1)}\big)$ has a uniform frequency grid (Fig.~\ref{fig:architecture}, vertical gray lines).
	The data~$d_I$ and PSD~$S_{\mathrm{n},I}$ are processed identically for each detector, yielding $3K$ segments $\big(d_I^{(k)}, S_{\mathrm{n},I}^{(k)}\big)$, each containing 16~frequency bins. 
	
	Each data segment is converted into a token embedding $ t\big(d_I^{(k)}, S_{\mathrm{n},I}^{(k)}, f^{(k)}, f^{(k+1)}, I\big)$ by a shared tokenizer network~$t$ (see Fig.~\ref{fig:architecture}, bottom left).
	Transformer-based architectures are invariant under token permutations, so conditioning tokens on frequency range and detector identity is necessary for interpreting each segment~\citep{dauphin_glu:2017}. 
	Finally, we append one additional learnable summary token to the sequence so that the model has a dedicated place to store relevant information extracted across tokens~\citep{devlin:2019, darcet_registers:2024}.
	
	The full token sequence is passed to a transformer encoder~\citep{vaswani:2017, wang:2019, xiong:2020}, which extracts information relevant for parameter estimation. After the final transformer layer, we compress the summary token to a 128-dimensional feature vector, which conditions a normalizing flow with a rational-quadratic spline coupling transform~\citep{durkan_nsf:2019}, using the same hyperparameters as in~\citep{dax_is:2022}.
	
	\sec{Training}We train the \dingoT model---the tokenizer, transformer encoder, and normalizing flow---end-to-end using a negative log likelihood loss. 
	To enable inference under missing data, we mask tokens randomly during training~\citep{cheng_mask2former:2021}. 
	This has the effect of dropping segments of data, so that the network learns to interpret inputs with missing data. 
	By applying a token mask~$m$ 
	to $(d, S_\mathrm{n})$, our loss takes the form
	\begin{equation}
		\mathcal{L} =  \mathbb{E}_{p(\theta)p(S_n)p(d|\theta, S_\mathrm{n})p(m)}\left[ - \log q\left(\theta| m(d), m(S_n)\right)\right],
	\end{equation}
	where $p(m)$ denotes the probability of sampling mask $m$.
	
	We employ a data-based approach to masking, i.e., we choose $p(m)$ to reflect realistic variations in data analysis settings. 
	This includes dropping all tokens corresponding to certain detectors, removing tokens to update minimum and maximum frequencies, and applying random cuts within the frequency range. 
	We also compare this approach to a masking strategy where tokens are dropped uniformly at random, but find that the data-based approach gives better performance. (See the Supplemental Material~\citep{supplemental} for complete details on masking.)
	
	We train on simulated data consisting of waveforms generated from the IMRPhenomXPHM model~\citep{pratten_imrphenomxphm:2021} with additive stationary Gaussian noise. 
	Our dataset includes $2.5\cdot 10^7$ intrinsic waveforms, with extrinsic parameters generated on-the-fly during training. 
	Priors are the same as \citep{dax_is:2022}, except for the luminosity distance, which we take as uniform over the range $[0.1, 6]~\mathrm{Gpc}$; we infer all parameters except for the phase, which is reconstructed in post-processing~\cite{dax_dingo:2021}. 
	We generate noise during training based on a collection of Welch PSDs estimated across O3~\citep{dax_dingo:2021}. 
	Models are trained on 8~NVIDIA-A100 GPUs using distributed data parallelization~\citep{loshchilov_adamw:2018, micikevicius_amp:2018} which takes approximately 9.5~days (183~epochs) for \dingoT.
	
	At inference time, we only include tokens corresponding to available data that we wish to analyze, omitting those associated with missing information (e.g., a non-observing detector or masked frequency ranges). 
	To validate and correct the resulting posterior, we apply importance sampling~(IS)~\citep{dax_is:2022}, reweighting samples from $q(\theta | d, S_n)$ to the exact posterior defined by the ranges of the original uniform-frequency data.
	Due to variance in the distribution of importance weights, the effective number of weighted samples is smaller than the number of unweighted samples. 
	This ratio (the sample efficiency) serves as a performance metric.
	
	\begin{figure}[t]
		\vspace*{.1cm}
		\subfloat{
			\stackinset{l}{0cm}{t}{-0.1cm}{\textbf{a~~Simulated events}}{
				\stackinset{l}{0cm}{t}{3.3cm}{\textbf{b~~Real data}}{
					\hspace*{-0.2cm}\includegraphics[width=\columnwidth]{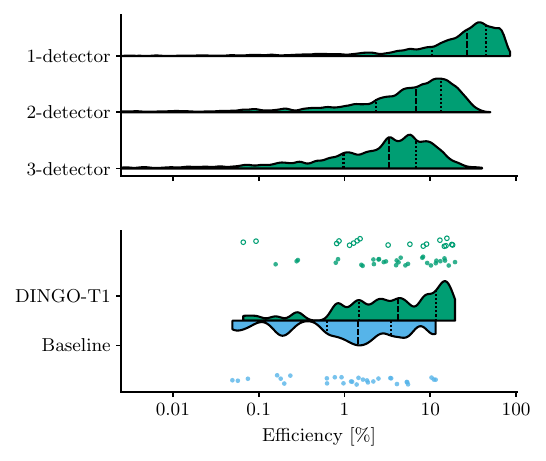}
				}
			}
		}
		\caption{(a)~Sample efficiency distribution of $1000$ simulated events per detector configuration for the \dingoT model, shown as violin plots.
			(b)~Sample efficiency distribution of \dingoT and the \dingo baseline across 48~real GW~events. Dots denote 3-detector events, while circles refer to 2-detector events which cannot be analyzed with the unflexible \dingo baseline. 
			The violin plots summarize the distributions, while the points for the individual events make the limited sample size explicit.
			The dashed line marks the median and the dotted lines the quartiles. 
			The efficiencies between simulated~(a) and real data~(b) are not directly comparable since the parameters are drawn from different distributions.
			\looseness=-1
		}
		\label{fig:efficiency_injections_events}
	\end{figure}
	
	\sec{Results}We first evaluate \dingoT on 1000~\emph{simulated} GW~signals for each subset of the three detectors~HLV, with injections performed over the full frequency range.
	For each injection, we generate $10^5$ posterior samples and perform~IS. 
	Initial sampling takes under five seconds, with phase reconstruction and IS requiring an additional five to ten minutes on 64~CPU cores. 
	Median efficiencies are 26.9\%, 6.8\%, and 3.3\% for \mbox{one-,} two-, and three-detector configurations, respectively (Fig.~\ref{fig:efficiency_injections_events}a). 
	Efficiencies decrease as more detectors are included, consistent with expectations: posteriors constrained by more detectors are narrower and therefore more challenging to learn for normalizing flows.
	To show that \dingoT is well calibrated, we provide \mbox{P--P~plots} for each detector configuration and additional information in the Supplemental Material~\citep{supplemental}.
	
	\begin{figure*}[t]
		\begin{minipage}[t]{.48\textwidth}
			\subfloat{
				\stackinset{l}{0cm}{t}{0cm}{\textbf{a}}{
					\stackinset{l}{3cm}{t}{0cm}{ GW190701\_203306}{
						\stackinset{l}{2.8cm}{t}{0.16cm}{%
							\includegraphics[width=0.28\textwidth]{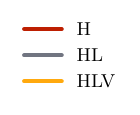}%
						}{
							\stackinset{l}{4.8cm}{t}{0.2cm}{%
								\includegraphics[width=0.4\textwidth]{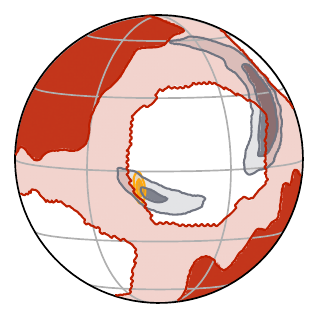}%
							}{
								\includegraphics[width=\textwidth]{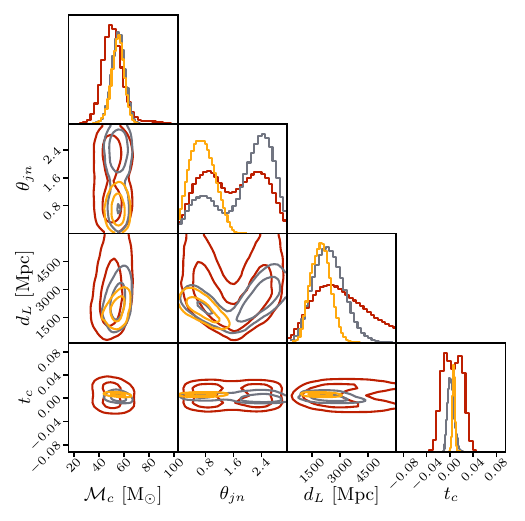}%
				}}}}
			}
		\end{minipage}
		\hfill
		\begin{minipage}[t]{.48\textwidth}
			\subfloat{
				\stackinset{l}{0.0cm}{t}{-0.2cm}{\textbf{b}}{
					\includegraphics[width=\textwidth]{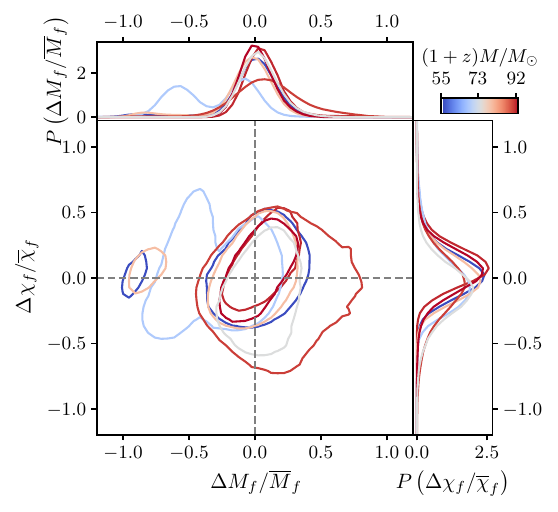}
				}
			}
		\end{minipage} 
		\caption{(a) Posterior distribution for \mbox{GW190701\_203306}, showing \dingoT analyses for three different detector configurations.
			(b) Inspiral-merger-ringdown consistency tests for seven events where the signal was analyzed with the \dingoT model on the inspiral and postinspiral part of the signal. The main panel shows the 90\% credible regions of the 2D posteriors on $(\Delta M_f / \overline{M}_f, \Delta \chi_f / \overline{\chi}_f)$ and the side panels show the marginalized posteriors.
			The color corresponds to the median redshifted total mass obtained from posterior samples of the full signal.
		} 
		\label{fig:results_flexible_model}
	\end{figure*}
	
	Second, we evaluate \dingoT on all BBHs from~O3 that fall within our prior, employing the specific data analysis settings from the official catalogs~\citep{ligo_gwtc-21:2024, ligo_gwtc-3:2023}.
	For each event, we obtain posterior samples as above and summarize sample efficiencies in Fig.~\ref{fig:efficiency_injections_events}b.
	Across the 48~events, we find a median efficiency of 4.2\%; 2-detector events perform slightly better at 4.5\%, while 3-detector events reach 4.2\%. 
	This outperforms a baseline \dingo NPE model trained for fixed data analysis settings, which obtains a median efficiency of 1.4\% across 30~HLV events evaluated on the full frequency range. 
	In general, sample efficiencies of real data (in contrast to injections) can be negatively impacted by misspecification of the signal and noise model.
	Excluding such events increases the median sample efficiency from 4.2\% to 5.2\% for all events (see Supplemental Material~\citep{supplemental}).
	
	It should be noted that our results are not directly comparable to past work~\cite{dax_is:2022} by some of us, which reported $\sim\!10\%$ median efficiency on O3~events. 
	That study performed HL~analyses only; when restricting to HL~detectors we find 11\% median efficiency with \dingoT (see Supplemental Material~\citep{supplemental}). 
	Ref.~\cite{dax_is:2022} also incorrectly accounted for the window factor in the likelihood, likely resulting in lower efficiencies~\cite{talbot_window_fix:2025}. 
	More importantly, Ref.~\cite{dax_is:2022} used \emph{group-equivariant} NPE (GNPE,~\cite{dax_dingo:2021,dax_gnpe:2021}), an algorithm which simplifies data representations by iteratively aligning merger times in each detector.
	GNPE improves accuracy, but results in substantially longer analysis times as it breaks access to the probability density. 
	\dingoT is based on standard NPE, making inference faster and simpler.
	\looseness=-1
	
	\sec{Applications}To explore the impact of detector configuration on the posterior, we evaluate every event on all possible combinations.
	While such studies are computationally infeasible with standard samplers, total inference times of under ten minutes per event make it a reasonable analysis with \dingoT. 
	We illustrate the effect on the posterior distribution for an example event in Fig.~\ref{fig:results_flexible_model}a.
	(See Supplemental Material~\citep{supplemental} for detailed discussions.)
	
	Finally, we utilize the flexibility of \dingoT to perform inspiral-merger-ringdown~(IMR) consistency tests probing general relativity~\citep{ghosh:2018, ligo_test_gr_gwtc2:2020, lvk_gwtc3_tests_gr:2025}.
	For these tests, the posteriors of the final mass~$M_f$ and spin~$\chi_f$ are independently estimated for the inspiral (low) and postinspiral (high frequency) part of the signal, separated by an event-specific cut-off frequency~$f_\mathrm{cut}$.
	We perform IMR~consistency tests on seven events from~O3 where we obtain high enough efficiency to draw 5,000 posterior samples---with all analyses carried out using the same \dingoT model.
	We display the fractional deviations between the inspiral and postinspiral parameters in Fig.~\ref{fig:results_flexible_model}b and find that our analyses show overall consistency with general relativity. 
	
	\sec{Conclusions}We introduced \dingoT, a transformer-based model that enables fast and flexible GW~parameter inference across varying data analysis settings, including detector configurations and frequency ranges. 
	We validated \dingoT on an array of simulated and real data sets, including a realistic catalog analysis involving 17~data configurations. 
	We argued that \dingoT can be used to probe systematics, and demonstrated it can be applied for IMR~consistency tests. 
	Thanks to improved architecture and training, we obtain increased median sample efficiency on O3 HLV events of 4.2\%, compared to 1.4\% for a NPE~baseline. 
	As we do not leverage knowledge of time-translation equivariances in \dingoT, the efficiencies do not fully reach GNPE-level performance.
	However, \dingoT has greatly increased flexibility and maintains access to the density, making it significantly more useful in practice and enabling further scaling and fine-tuning studies.
	
	A natural next step would be to consider variations in signal duration by conditioning tokens on frequency resolution information; however, this would require substantial modifications to the current training pipeline and is therefore left to future work.
	Another promising direction is to explore alternative data representations such as time domain or time-frequency domain~\cite{Cornish:2020odn} (employing vision transformers~\citep{dosovitskiy:2021}), where simpler signal morphology may lead to improved training. 
	In the context of LISA, our approach offers a natural solution to the problem of expected data gaps~\citep{burke:2025}. 
	In general, the proposed transformer framework and training strategy can be readily extended and adapted to other tasks like signal detection and the \dingoT transformer encoder could serve as a general-purpose GW~compression backbone. 
	By concentrating on adaptable models that can be shared across tasks, computational effort can thereby be invested where it most benefits scaling and performance.
	
	\vspace{0.5em}
	This work has made use of many open-source Python packages,
	including 
	\texttt{astropy}~\citep{astropy:2022},
	\texttt{BayesWave}~\citep{cornish:2014, littenberg:2014, cornish:2020},
	\texttt{Bilby}~\citep{bilby}, 
	\texttt{bilby\_pipe}~\citep{bilby_pipe}, 
	\texttt{corner}~\citep{corner:2016}, 
	\texttt{glasflow}~\citep{glasflow},
	\texttt{LALSuite}~\citep{lalsuite, swiglal}, 
	\texttt{ligo.skymap}~\citep{ligo_skymap},
	\texttt{matplotlib}~\citep{matplotlib:2007}, 
	\texttt{nflows}~\citep{nflows}, 
	\texttt{numpy}~\citep{numpy:2020}, 
	\texttt{pandas}~\citep{pandas},
	\texttt{PyCBC}~\citep{pycbc}, 
	\texttt{PyTorch}~\citep{pytorch}, 
	\texttt{seaborn}~\citep{seaborn}, and
	\texttt{wandb}~\citep{wandb}. 
	For training, we employ the AdamW~\citep{loshchilov_adamw:2018} optimizer and automatic mixed precision~\citep{micikevicius_amp:2018}. 
	The accessible color schemes in
	our figures are based on~\citep{colorblind_tol, colorblind_okabe_ito}.
	
	\vspace{0.5em}
	
	\begin{acknowledgments}
		\sec{Acknowledgments}We thank Michael P\"urrer and Vincent Berenz for helpful discussions, and Ik Siong Heng for comments on the manuscript.
		We thank the reviewers for their detailed questions and comments on the draft.
		A.K. thanks the International Max Planck Research School for Intelligent Systems~(IMPRS-IS) for support. S.R.G. is supported by a UKRI Future Leaders Fellowship (grant number MR/Y018060/1).
		A.B.'s research is supported in part by the European Research Council~(ERC) Horizon Synergy Grant “Making Sense of the Unexpected in the Gravitational-Wave Sky” grant agreement no. GWSky–101167314.
		The computational work for this \emph{Letter} was performed on the Atlas cluster at the Max Planck Institute for Intelligent Systems. 
		This research has made use of data or software obtained from the Gravitational Wave Open Science Center (gwosc.org), a service of the LIGO Scientific Collaboration, the Virgo Collaboration, and KAGRA. This material is based upon work supported by NSF's LIGO Laboratory which is a major facility fully funded by the National Science Foundation, as well as the Science and Technology Facilities Council~(STFC) of the United Kingdom, the Max-Planck-Society~(MPS), and the State of Niedersachsen/Germany for support of the construction of Advanced LIGO and construction and operation of the GEO600 detector. Additional support for Advanced LIGO was provided by the Australian Research Council. Virgo is funded through the European Gravitational Observatory~(EGO), by the French Centre National de Recherche Scientifique~(CNRS), the Italian Istituto Nazionale di Fisica Nucleare~(INFN) and the Dutch Nikhef, with contributions by institutions from Belgium, Germany, Greece, Hungary, Ireland, Japan, Monaco, Poland, Portugal, Spain. KAGRA is supported by Ministry of Education, Culture, Sports, Science and Technology~(MEXT), Japan Society for the Promotion of Science~(JSPS) in Japan; National Research Foundation~(NRF) and Ministry of Science and ICT~(MSIT) in Korea; Academia Sinica (AS) and National Science and Technology Council~(NSTC) in Taiwan.
	\end{acknowledgments}
	
	\vspace{0.5em}
	\sec{Data availability}The instructions to reproduce the findings of this article are openly available at \href{https://github.com/dingo-gw/dingo-T1}{https://github.com/dingo-gw/dingo-T1} which includes the code, the settings files, as well as information about downloading and performing inference with the neural networks~\citep{kofler_dingoT_zenodo} which were used to generate the reported results.
	
	\newpage
	\clearpage

\begin{center}
{\large\bfseries Supplemental Material}
\end{center}

\section{Full amortization in gravitational wave posterior estimation}
In gravitational wave~(GW) inference, the data representation depends on the analysis settings, such as available detectors and frequency ranges. 
We define full amortization in this context as the ability to perform parameter estimation with a single model that generalizes across these settings. 
\dingoT advances toward this goal by learning a flexible representation that adapts to varying data configurations at inference time.

Formally, the goal of GW inference is to estimate the Bayesian posterior over parameters~$\theta$
\begin{equation}\label{eq:app-Bayes}
    p(\theta | d) = \frac{p(\theta)p(d | \theta)}{p(d)}~,
\end{equation}
defined by prior $p(\theta)$ and likelihood $p(d|\theta)$. 
Assuming additive, stationary, and Gaussian detector noise characterized by its noise amplitude---the power spectral density~(PSD)~$S_n(f)$---the likelihood takes the form
\begin{equation}
    p(d|\theta, S_n) \propto \exp \left( - \frac{1}{2} \big(d-h(\theta) | d- h(\theta)\big)_{S_n} \right)~,
\end{equation}
where $h(\theta)$ denotes the simulated GW signal for source parameters~$\theta$.
With both $d$ and $h(\theta)$ represented as frequency series, the noise-weighted inner product is given by
\begin{equation}\label{eq:app-inner-product}
    \big( a | b \big)_{S_n} = 4 \sum_I \mathcal{R} \int_{f_{\min, I}}^{f_{\max, I}} \frac{a_I^*(f) b_I (f)}{S_{n, I}(f)} ~\mathrm{d}f~.
\end{equation}
This inner product---and consequently the likelihood---quantifies how well the simulated signal~$h(\theta)$ matches the observed data~$d$ across detectors~$I$.
Although the posterior~$p(\theta | d)$ is often written as depending only on~$d$, it in fact reflects several modeling and analysis choices implicit in Eqs.~\eqref{eq:app-Bayes}--\eqref{eq:app-inner-product}: the prior distribution, the waveform model used to generate $h(\theta)$, the noise PSDs, the set of detectors, the frequency range~$[f_{\mathrm{min}, I}, f_{\mathrm{max}, I}]$, and the discretization $\Delta f$ defining the resolution of the integral in the inner product. 
These quantities can vary between events, meaning that a fully amortized inference model~$q(\theta|d)$ must accommodate such variations.
The \dingo framework already supports changing PSDs via PSD-conditional models~$q(\theta|d,S_n)$~\cite{dax_dingo:2021}, and can, in restricted cases, also accommodate varying priors~\cite{dax_bns:2025}.
Adjustments to the lower frequency cutoff~$f_{\mathrm{min},I}$ have been explored by masking the corresponding data segments with zeros~\cite{dax_bns:2025}, though this approach can lead to suboptimal performance and biases in the posterior estimates~\citep{wang_missing_data:2024, verma:2025}.

\begin{figure}[t]
    \includegraphics[width=\columnwidth]{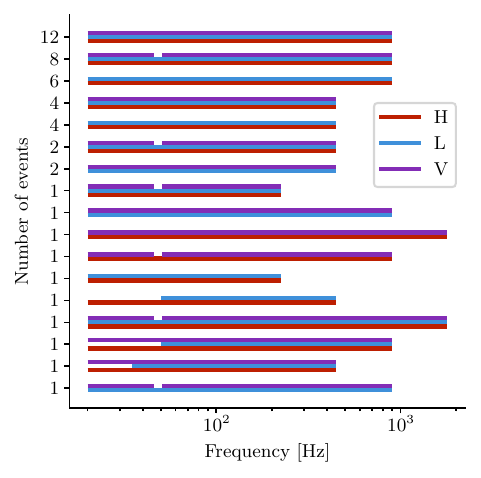}
    \caption{The 48 GW events considered in this study are analyzed in 17 different data analysis settings in O3~LVK catalogs~\cite{ligo_gwtc-21:2024,ligo_gwtc-3:2023}. 
    Each event is based on data from a subset of the three detectors~(HLV) and spans frequency ranges that vary with data quality and source properties. The \dingoT model accommodates all of these settings with a single neural network.
    } 
    \label{fig:settings_o3}
\end{figure}

A flexible model like \dingoT provides the key to achieving full amortization in GW~inference since such a model can flexibly accommodate the diverse data analysis settings encountered in practice, as illustrated in Fig.~\ref{fig:settings_o3} for O3~BBH~events.
The detector configuration can change from event to event due to maintenance, instrumental issues, earthquakes, or poor data quality~\citep{ligo_gwtc-21:2024,ligo_gwtc-3:2023}.
Similarly, the analysis frequency range may be adjusted: the minimum frequency, typically $f_\mathrm{min} = \SI{20}{\hertz}$, can be increased to exclude non-Gaussian noise artifacts that cannot be subtracted from the data~\citep{ligo_gwtc-21:2024}, as is the case for the three events highlighted in Tab.~\ref{table:event_configs}.
The maximum frequency of each event is determined by the sampling rate $f_s$ needed to fully resolve the signal, with $f_\mathrm{max} \propto f_s/2$~\citep{ligo_gwtc-21:2024, ligo_gwtc-3:2023}.
In addition to such global adjustments, narrow frequency bands may be excluded to correct for systematic calibration errors, a procedure known as ``PSD notching.''
For example, several O3b~events involving the Virgo detector exhibited a narrow-band calibration error between \SI{46}{\hertz} and \SI{51}{\hertz} due to a mis-specification in the calibration models.
To mitigate its impact on parameter estimation, this frequency interval is effectively removed in standard samplers by assigning an artificially large noise level~\citep{ligo_gwtc-3:2023}.

In summary, \dingoT amortizes over key data-analysis variations---noise PSDs, detector setups, frequency ranges, and calibration errors---while full amortization over frequency resolution, priors, and waveform models remains future work.
\looseness=-1

\section{Data settings}
\sec{Prior}
The prior ranges are listed in Tab.~\ref{table:priors} and follow~\citep{dax_is:2022}, except for the luminosity distance, for which we adopt $d_L \in [0.1, 6]~\mathrm{Gpc}$.
This choice reflects that we train a single \dingoT model rather than three separate models for different $d_L$ ranges.
\begin{table}[ht]
    \caption{Training priors of \dingo and \dingoT models.}
    \label{table:priors}
    \centering
    \begin{tabularx}{\columnwidth}{l c}
        \toprule
        Parameter & Prior \\
        \midrule
        \multicolumn{2}{l}{\textit{Intrinsic parameters}} \\
        Chirp mass~$\mathcal{M}_c$ & $\mathcal{U}[15, 150]~M_\odot$ \\
        Mass ratio~$q$ & $\mathcal{U}[0.125, 1]$ \\
        Spin magnitudes~$a_1, a_2$ & $\mathcal{U}[0, 0.99]$ \\
        Tilt angles~$\theta_1, \theta_2$ & $\mathrm{Sine}(0,\pi)$ \\
        Azimuthal spin angle~$\phi_{12}$ & $\mathcal{U}[0, 2\pi)$ \\
        Spin-orbit azimuth~$\phi_\mathrm{jl}$ & $\mathcal{U}[0, 2\pi)$ \\
        Inclination~$\theta_\mathrm{JN}$ & $\mathrm{Sine}(0,\pi)$ \\
        Coalescence phase~$\phi$ & $\mathcal{U}[0, 2\pi)$ \\
        Component masses $m_1, m_2$ (constraint) & $[10, 120]~M_\odot$\\
        \midrule
        \multicolumn{2}{l}{\textit{Extrinsic parameters}} \\
        Right ascension~$\alpha$ & $\mathcal{U}[0, 2\pi)$ \\
        Declination~$\delta$ & $\mathrm{Cosine}(-\pi/2, \pi/2)$ \\
        Polarization~$\psi$ & $\mathcal{U}[0, \pi)$ \\
        Time at geocenter~$t_c$ & $\mathcal{U}[-0.1, 0.1]~\si{\second}$ \\
        Luminosity distance~$d_L$ & $\mathcal{U}[0.1, 6]~\si{\giga \parsec}$ \\
        \bottomrule
    \end{tabularx}
\end{table}

\sec{Frequency settings} 
For the 48~events consistent with the chosen prior ranges, we extracted the frequency ranges from the \mbox{GWTC-2.1}~\citep{ligo_data_release_gwtc21_zenodo:2022} and \mbox{GWTC-3}~\citep{ligo_data_release_gwtc3_zenodo:2021} data releases and the detector configurations from the corresponding catalogs~\citep{ligo_gwtc-21:2024, ligo_gwtc-3:2023}.
These configurations cover an overall frequency range of $[20,~1792]~\si{\hertz}$, and all events are analyzed using a fixed signal duration of \mbox{$T = \SI{8}{\second}$}, leaving amortization over different frequency resolutions for future work. 
A complete summary of the event-specific settings is provided in Tab.~\ref{table:event_configs}.

\sec{Multibanding} 
We employ multibanding~\citep{vinciguerra_mb:2017, morisaki:2021} to compress GW signals without loss of information before passing them to the \dingoT model. 
Multibanding exploits the chirping structure of compact binary waveforms: As the frequency increases, the time remaining until merger~$t(f)$ decreases, so the signal occupies a progressively shorter duration in the time domain. 
Consequently, the waveform becomes increasingly oversampled in the frequency domain at high frequencies.
To avoid redundant sampling, neighboring frequency bins can be averaged where the waveform varies slowly over frequency, resulting in a non-uniform frequency grid with finer resolution at low frequencies and coarser resolution at high frequencies.

Following~\citep{dax_bns:2025}, the frequency resolution is typically decreased by a factor of two from one multibanded frequency band to the next going from low to high frequencies, with band boundaries (called "nodes") determined by requiring 32~frequency bins per waveform period. 
However, this approach of determining the bands is not directly applicable to precessing waveforms such as IMRPhenomXPHM~\citep{pratten_imrphenomxphm:2021}, because precession introduces complex behavior between zero-crossings. 
To address this, we decimate $10^3$~whitened waveforms for a range of potential compression factors and compute the difference between each decimated waveform and its high-resolution counterpart. 
This provides a local quality metric for information loss at each frequency. 
By constraining the maximally allowed local deviation, we identify the frequency above which a lower resolution can be used. 
Going from lower to higher frequencies, we group consecutive token segments of fixed size until resolution reduction is permitted, defining the nodes of the multibanded frequency domain.
To validate the compression as loss-free, we compute the mismatch between $10^3$~decimated waveforms and their interpolated counterparts under extreme conditions (minimal chirp mass and maximal time-shift). 
We obtain a maximal mismatch of \mbox{$1.3\cdot 10^{-3}$}, comparable to the lower end of mismatches reported between IMRPhenomXPHM and numerical relativity simulations~\citep{pratten_imrphenomxphm:2021}.

The resulting multibanded domain defined on the frequency range~$[20,~ 1810]~\si{\hertz}$ achieves a compression factor of~$\sim 9$ 
and is used to generate a dataset of $2.5 \cdot 10^7$ waveforms for training.
Since the multibanding nodes are defined based on the size of the token segment, we can directly partition the multibanded data into segments of fixed length that serve as tokenizer input.
We adopt a token size of 16: smaller tokens (4~or~8) increase sequence length and training time without yielding noticeable improvements in model performance during training. 
Larger token sizes would result in less fine-grained control over frequency ranges at inference time.
We include the specific frequency ranges of the transformer tokens employed in this work in Tab.~\ref{tab:frequencies_for_tokens}.

\newcommand{\thinmidrule}{\specialrule{0.3pt}{0pt}{0pt}}
\begin{table}[ht]
    \caption{Frequency ranges corresponding to tokens used in \dingoT. All numbers in~\si{Hz}.}
    \label{tab:frequencies_for_tokens}
    \centering
    \setlength{\tabcolsep}{6pt}
    \begin{tabular}{
        S 
        S 
        S[table-format=4.4] 
        S[table-format=4.4]
        }
        \toprule
         {Nodes of }  & {\multirow{3}{*}{{Resolution}}} & {\multirow{3}{*}{$f_\mathrm{min}^{(k)}$}} & {\multirow{3}{*}{$f_\mathrm{max}^{(k)}$}}\\
         {multibanded} & && \\
         {domain} & & & \\
        \midrule
        20.0 & 2 & 20.0 & 21.875\\
         & & 22.0 & 23.875\\
         & & {\vdots} & {\vdots} \\
         & & 36.0 & 37.875 \\ \thinmidrule
        38.0 & 4 & 38.0625 & 41.8125\\
         & & 42.0625 & 45.8125\\
         & & 46.0625 & 49.8125 \\ \thinmidrule 
        50.0 & 8 & 50.1875 & 57.6875\\
         & & 58.1875 & 65.6875 \\ \thinmidrule
        66.0 & 16 & 66.4375 & 81.4375 \\ \thinmidrule
        82.0 & 32 & 82.9375 & 112.9375\\
         & & 114.9375 & 144.9375\\
         & & {\vdots} & {\vdots} \\
         & & 1778.9375 & 1808.9375\\
        \bottomrule
    \end{tabular}
\end{table}

\section{Architecture}
Each low-dimensional segment of 16~multibanded frequency bins is mapped to a token embedding vector of length~$d_\mathrm{model} = 1024$ by the tokenizer, using fully connected layers together with a 512-dimensional residual block. 
Conditional inputs---including the segment boundaries $\big(f^{(k)}, f^{(k+1)}\big)$ and the detector identity $I$ (encoded as a one-hot vector)---are injected within the residual block via a gated linear unit~\citep{dauphin_glu:2017}. 
This is illustrated on the left of Fig.~1 in the main text at the example of multibanded data from the Hanford detector.
We also experimented with an unconditional tokenizer together with additive positional encodings as in standard language models~\citep{vaswani:2017}. 
However, such encodings are less suitable for continuous, non-uniform quantities like frequency values, and multiple encodings would be needed to incorporate both $\big(f^{(k)}, f^{(k+1)}\big)$ and~$I$, resulting in a large number of design choices.
Over the full frequency range, we obtain 69~tokens per detector. 
Tokens from all detectors are concatenated together with a learnable, randomly initialized summary token~\citep{devlin:2019, darcet_registers:2024}, yielding a total of $n = 69 \cdot 3 + 1 = 208$ tokens, each of size~$d_\mathrm{model}$.
These token embeddings, represented as \mbox{$X \in \mathbb{R}^{n \times d_\mathrm{model}}$}, serve as the input to the transformer encoder which captures complex correlations between the $n$~tokens via masked self-attention:
\begin{equation}
    \mathrm{Attention}(Q, K, V) = \mathrm{softmax} \left( \frac{QK^T}{\sqrt{d_k}} + M\right) V
\end{equation}
where the mask is defined as $M_{ij} = 0$ for allowed and $M_{ij} = - \infty$ for disallowed positions~\citep{vaswani:2017, cheng_mask2former:2021} (see Fig.~1 of the main text on the right). 
The query, key, and value matrices are computed as
\begin{equation}
    Q = X \cdot W^Q,~ K = X \cdot W^K,~ V = X \cdot W^V 
\end{equation} 
with weights $W^K, W^Q, W^V \in \mathbb{R}^{d_\mathrm{model} \times d_k}$.
To capture multiple representation subspaces, attention is extended to $h$~heads, whose outputs are concatenated and combined with the input~$X$ via a residual connection. 
Each token is passed through a feed-forward layer~\citep{vaswani:2017}, with pre-layer normalization~(LN) applied before both the self-attention and feed-forward blocks~\citep{wang:2019, xiong:2020}. 
We adopt the pre-LN variant instead of the original post-LN design~\citep{vaswani:2017} as it enables stable training without learning rate warm-up. 
After $N=8$~transformer layers with $h=16$~attention heads and feed-forward hidden size~2048, the summary token is extracted and projected to a 128-dimensional context vector using a linear layer. 
We found that varying the context size between 56~and 256~dimensions does not significantly affect the training loss, indicating that 128~values suffice for subsequent posterior estimation.

For density estimation, we employ a conditional normalizing flow based on the rational-quadratic spline coupling transform~\citep{durkan_nsf:2019}, using the same architecture as in~\citep{dax_is:2022}. 
The only difference is that we train a standard NPE network and do not include group equivariances between parameters and data~\citep{dax_gnpe:2021}.
The conditional normalizing flow architecture is identical for all models.

We compare \dingoT to a standard \dingo~NPE model, which uses a residual network similar to~\citep{dax_dingo:2021} and employs layer norm instead of batch norm in the embedding network. 
The initial layer is seeded with components of a singular value decomposition, but we do not keep the initial embedding layer fixed~\citep{dax_dingo:2021} such that we can utilize the same optimizer and scheduler as the \dingoT models. 
A summary of model sizes is provided in Tab.~\ref{table:model_sizes}.

\begin{table}[ht]
    \caption{Model sizes of the compared architectures.}
    \label{table:model_sizes}
    \centering
    \setlength{\tabcolsep}{6pt}
    \begin{tabularx}{\columnwidth}{l c c c}
        \toprule
         \multirow{3}{*}{Method} & \multicolumn{3}{c}{Number of parameters}\\
         \cmidrule{2-4}
         & Embedding & Normalizing & \multirow{2}{*}{Total}\\
        & network & flow &\\
        \midrule
        \dingo Baseline & 22 Mio. & 92 Mio. & 114 Mio.\\
        \dingoT & 68 Mio. & 92 Mio. & 160 Mio. \\
        \bottomrule
    \end{tabularx}
\end{table}

\section{Masking strategies during training}
\begin{figure*}[ht]
    \subfloat{
        \stackinset{l}{0.1cm}{t}{0cm}{\textbf{a}}{
            \stackinset{l}{8.9cm}{t}{0cm}{\textbf{b}}{
            \includegraphics[width=0.95\textwidth]{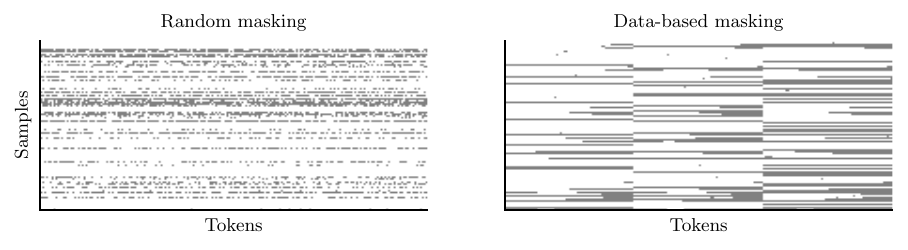}
            }
        }
    }
    \caption{\label{fig:masking_illustration}
    Comparison of (a)~random and (b)~data-based masking across 100~samples.
    Data-based masking jointly removes tokens from the same detector, yielding structured patterns, while random masking produces unstructured, scattered masks.
    } 
\end{figure*}

To ensure the model can flexibly handle different data-analysis settings at inference time, it must be trained on data that reflect such variability.
We therefore adopt a training strategy that mimics changing analysis conditions by removing tokens during training.
Specifically, we compare two masking strategies: one uninformed by token relationships (random masking), and one that incorporates data-based structure, such as jointly masking tokens from the same detector (data-based masking), illustrated in Fig.~\ref{fig:masking_illustration}.

\sec{Random masking}
In the naive masking strategy, 40\% of training samples are partially masked.
For each masked sample, we draw the number of tokens to remove as
$n_\mathrm{masked} \sim \mathcal{U}[0,~181]$ where the upper bound allows for single-detector scenarios with additional frequency masking.
The token mask~$m$ is created by randomly selecting $n_\mathrm{masked}$ tokens, without assuming any domain knowledge.
This strategy thus provides unstructured masks (Fig.~\ref{fig:masking_illustration}a), serving as a simple baseline for testing the model’s robustness to missing information.

\sec{Data-based masking}
Informed by the structure of GW~data, we design a second masking strategy that explicitly captures realistic variations in detector configurations and frequency ranges.
This procedure accounts for
(1)~missing detectors,
(2)~changes in the frequency range, and
(3)~narrow-band frequency removal (“PSD notching”).

(1)~To simulate missing detectors, we randomly decide whether to mask none (60\%), one (30\%), or two (10\%) detectors.
Among the available detectors (H/L/V), the probabilities of masking are 30\%, 30\%, and 40\%, respectively, reflecting Virgo’s larger downtime relative to the LIGO instruments~\citep{ligo_gwtc-3:2023}.
(2)~To model changing frequency ranges, we apply frequency updates to 25\% of training samples.
For these, we mask the lower part (10\%), upper part (70\%), or both (20\%) of the range, inspired by Fig.~\ref{fig:settings_o3}.
Since frequency updates often affect all detectors simultaneously, we apply the same mask across all detectors in 70\% of these cases.
New cutoff frequencies are drawn from $f_\mathrm{min,new} \sim \mathcal{U}[20,180]~\mathrm{Hz}$ and $f_\mathrm{max,new} \sim \mathcal{U}[80,~1810]~\mathrm{Hz}$ to accommodate frequency updates required for inspiral-merger-ringdown consistency tests.
In cases of $f_\mathrm{min,new} > f_\mathrm{max,new}$, we randomly decide to either mask the lower or upper range, depending on the probabilities given above.
If a boundary falls within a token, the entire token is masked, ensuring that all affected frequency regions are consistently removed.
(3)~To emulate PSD notching, we mask a narrow frequency band in 10\% of training samples, where the lower bound is sampled uniformly across the frequency range and the width is drawn as $\Delta f_\mathrm{mask} \sim \mathcal{U}[0,~10]~\mathrm{Hz}$.
Overall, this data-based masking procedure closely reflects the detector and frequency conditions encountered in real gravitational-wave analyses, resulting in clearly visible boundaries between tokens from one detector in Fig.~\ref{fig:masking_illustration}b.
On average, 18.0\% of tokens are masked during training for random masking and 25.5\% for data-based masking.
We experimented with masking percentages ranging from 15\% to 30\%: Low masking percentages resulted in lower performance in two-detector events, but applying masks to a large number of samples during training can lead to decreased performance in three-detector events.

\section{Training}
We train all models using distributed data-parallel multi-GPU training with automatic mixed precision~\citep{micikevicius_amp:2018} to reduce memory usage.
Training is performed on 8~NVIDIA~A100-SXM4-80GB~GPUs (CUDA~12.1) using the AdamW~\citep{loshchilov_adamw:2018} optimizer ($\beta_1 = 0.8$, $\beta_2 = 0.99$) with a learning rate of~0.001, weight decay of~0.005, and a \texttt{ReduceLROnPlateau} scheduler that halves the learning rate when the validation loss stagnates for 10~epochs.
To ensure comparability, all models are trained with a fixed batch size of~16{,}384.
Due to memory constraints, the \dingoT models require two gradient accumulation steps per optimizer update, effectively doubling their training time.
The total training duration further depends on the scheduler convergence and early-stopping conditions; detailed timings and epoch counts are summarized in Tab.~\ref{table:trainig_times}.
Because of the long training times, we do not perform extensive hyperparameter optimization beyond a small scan over the initial learning rate, as well as $\beta_1$ and $\beta_2$ considering the first 50 epochs.

\begin{table}[ht]
    \caption{
    Training times determined by early stopping.}
    \label{table:trainig_times}
    \centering
    \setlength{\tabcolsep}{6pt}
    \begin{tabularx}{0.88\columnwidth}{l c r@{}l @{\,} r@{}l}
        \toprule
        Method & Epochs\hspace{+0.5cm} & \multicolumn{4}{c}{\hspace{-0.5cm}Training time}\\
        \midrule
        \dingo Baseline & 273\hspace{+0.5cm} & 3&d& 1&h\\
        \dingoT Random & 219\hspace{+0.5cm} & 11&d& 17&h\\
        \dingoT Data-based & 183\hspace{+0.5cm} & 9&d& 9&h\\
        \bottomrule
    \end{tabularx}
\end{table}

\section{Inference}
During inference, the model has to adapt to the data analysis settings of the event of interest. 
Since tokens are treated as indivisible elements, any token overlapping a masked range is completely removed, which can lead to slightly larger effective exclusions than strictly necessary.
Potential effects on~$q(\theta | d)$ are mitigated by performing importance sampling~(IS, \citep{dax_is:2022}) in the uniform frequency domain.
In IS, importance weights~$w_i = p(\theta_i | d, S_{n}) p(\theta_i) / q(\theta_i | d, S_{n})$ are computed for each sample~$\theta_i \sim q(\theta_i | d)$. 
For $N$~weighted samples, the number of effective samples from the posterior~$p(\theta | d)$ corresponds to $N_\mathrm{eff} = (\sum_{i = 1}^N w_i )^2 / (\sum_{i = 1}^N w_i^2)$ and we employ the sample efficiency $\epsilon = N_\mathrm{eff} / N$ as a performance metric.
In practice, a specific number of effective samples is required to obtain a reliable posterior estimate, with $N_\mathrm{eff} = 5,000$ defined by the LVK collaboration.
Efficiencies $\epsilon \gtrsim 1\%$ therefore indicate that this requirement can be reached with minimal computational cost during IS; higher efficiencies (e.g., 2\% vs.~15\%) do not substantially reduce runtime and are therefore less critical.
For $\epsilon < 1 \%$, it is still possible to obtain a sufficient number of effective samples, but at an increased computational cost.

In principle, it is possible to further improve performance of \dingoT by incorporating additional knowledge about equivariances between data and parameters, called \textit{group-equivariant} NPE~(GNPE~\citep{dax_gnpe:2021,dax_gnpe:2021}).
However, exploiting these equivariances has two main drawbacks:
(i)~GNPE relies on the convergence of a costly iterative Gibbs sampling procedure at inference time since two models jointly estimate the data standardization and posterior distribution. 
(ii)~GNPE only provides samples from the posterior distribution and the posterior density has to be recovered by training an unconditional normalizing flow to facilitate~IS.
These points result in a complicated and impractical inference procedure.
Furthermore, large inference times of roughly~\SI{1}{\hour} for $10^5$~importance-sampled posterior samples on one NVIDIA-A100 GPU and 64~CPUs~\citep{dax_is:2022} limit exploratory interaction when adapting data analysis settings. 
In contrast, inference times of the \dingoT models are on the order of five to ten minutes for the same number of posterior samples, with initial low-latency samples being available within five seconds.
Inference times of \dingoT are dominated by analytically reconstructing the phase~$\phi_c$ which suffers from an inefficient implementation for IMRPhenomXPHM~\citep{dax_is:2022}.

\section{Comparing masking strategies}
To compare the performance of the two masking strategies, we evaluate both models---trained with either random or data-based masking---on simulated and real events.

\begin{figure}[ht]
    \centering
    \includegraphics[width=\columnwidth]{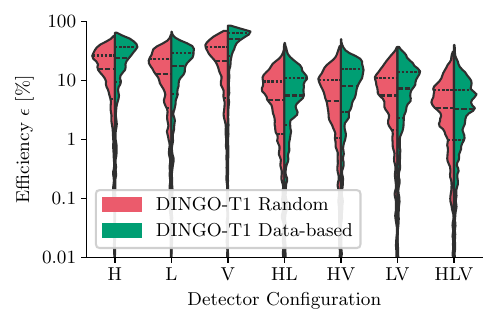}
    \caption{
    Sample efficiency distributions for 1000 simulated GW~signals evaluated with different detector configurations, shown as violin plots. The dashed line represents the median, the dotted lines the quartiles. We compare data-based masking with random token masking for the same \dingoT architecture.}
    \label{fig:injections_violin_plot}
\end{figure}

\sec{Simulated data}For 1000~simulated signals, we draw $10^5$~posterior samples from each \dingoT model across all detector configurations evaluated on the full frequency range and show the obtained sample efficiencies in Fig.~\ref{fig:injections_violin_plot}.
While the model trained with random masking achieves median efficiencies of 15.7\% (1-detector), 4.9\%~(2-detector), and 3.4\%~(3-detector), the efficiencies of the model trained with data-based masking are 26.9\%, 6.8\%, and 3.3\% for one-, two-, and three-detector configurations, respectively. 
By only changing the detector configuration for each injection, we observe that the model trained with random masking performs worse on 1-~and 2-detector settings.
This suggests that randomly masking tokens during training does not translate into performance equivalent to data-based masking.
Since we do not apply any frequency masking in this experiment, the random masking seems to have a small advantage on HLV~injections.
However, the frequency range has to be adapted for observed signals.

\begin{figure}[ht]
    \includegraphics[width=\columnwidth]{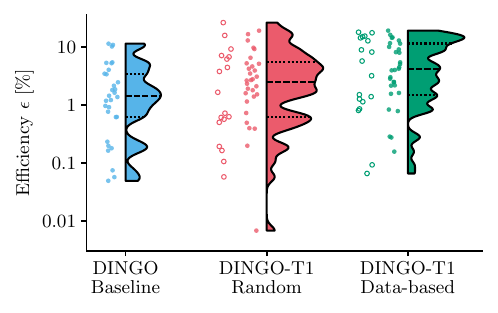}
    \caption{
    Sample efficiency distributions of \dingo models across the 48~real GW~events, shown as violin plots. Dots denote 3-detector events, while circles refer to 2-detector events which cannot be analyzed with the fixed \dingo baseline model. 
    The violin plots summarize the distributions, while the points for the individual events make the limited sample size explicit.
    The dashed line represents the median, the dotted lines the quartiles. 
    For the \dingo baseline, we only include the 30~events for which data is available in all three detectors.
    The individual efficiencies for all events are included in Tab.~\ref{table:efficiencies_events}.
    }
    \label{fig:events_violin_plot}
\end{figure}

\sec{Real events}We compare the impact of data-based and random masking on 48~out of a total of 84~observed BBH signals from O3~\citep{ligo_gwtc-21:2024, ligo_gwtc-3:2023} compatible with the chosen prior ranges which are analyzed with variable data settings (see Fig.~\ref{fig:settings_o3}).
We summarize the obtained sample efficiencies in Fig.~\ref{fig:events_violin_plot} and provide a detailed event list in Tab.~\ref{table:efficiencies_events}.
Across the 48~events, we find an overall median efficiency of 2.5\% for the model trained with random masking and 4.2\% for data-based masking.
Specifically, the model trained with random masking obtains 1.2\% for 2-detector events (HL: 5.4\%, HV: 0.3\%, LV: 0.7\%), and 2.7\% for 3-detector events.
\dingoT trained with data-based masking achieves median efficiencies of 4.5\% for 2-detector events (HL: 11.0\%, HV: 1.1\%, LV: 2.2\%), and 4.2\% for 3-detector events.
The baseline \dingo NPE~model reaches a median sample efficiency of~1.4\% across 30~HLV events evaluated on the full frequency range.
In Tab.~\ref{table:efficiencies_events}, we observe that sample efficiencies are consistent across models, meaning that a certain event with low efficiency in \dingoT with data-based masking also has similar or even worse performance in the model trained with random masking or in the \dingo NPE baseline.

Since data-based masking leads to better results than random masking on real events, we focus on the model with data-based masking in the following and refer to it as the \dingoT model.

\section{Impact of glitches on Performance}
\begin{figure}[ht]
    \includegraphics[width=\columnwidth]{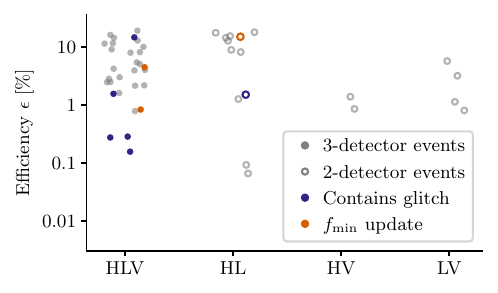}
    \caption{
    Sample efficiency distributions of \dingoT model with data-based masking across the 48~real GW~events where events affected by glitches are highlighted in different colors. 
    We distinguish events where deglitching was performed by updating~$f_\mathrm{min}$ (orange) since we take this into account in our analysis with \dingoT. 
    Events that would require other glitch mitigation strategies that we cannot address are shown in blue.
    Dots denote 3-detector events, while circles refer to 2-detector events. 
    }
    \label{fig:eff_events_glitches}
\end{figure}

Since we include events in our analysis that are subject to glitch mitigation in the official LVK analyses~\citep{ligo_gwtc-21:2024, ligo_gwtc-3:2023}, we investigate how model misspecification due to non-stationary noise affects the performance of \dingoT. 
To understand whether events with low sampling efficiency correspond to events where deglitching was applied, we specifically highlight these events in Fig.~\ref{fig:eff_events_glitches}.
While low efficiencies for several events can be attributed to glitches consistent with previous studies~\citep{dax_is:2022}, this does not hold for all events.
Additionally, events with an~$f_\mathrm{min}$ update show higher efficiencies consistent with the assumption that $f_\mathrm{min}$ is changed to mitigate non-Gaussian noise artifacts.

Furthermore, we report the median sample efficiency for each model for the full set of 48~events as well as for the 39~events where all glitch-affected events are excluded in Tab.~\ref{tab:eff_without_glitch_events}.
We observe that the efficiencies improve up to $\Delta \epsilon = + 1\%$ for \dingo-T1 trained with data-based masking when removing glitch-contaminated events.

\begin{table}[ht]
    \caption{Median efficiencies with and without glitch-affected events. For the latter, all events subject to glitch mitigation (including $f_\mathrm{min}$ update) in the official LVK analysis are removed.}
    \label{tab:eff_without_glitch_events}
    \begin{tabularx}{\columnwidth}{l c c c}
        \toprule
         \multirow{2}{*}{Median efficiency}
         & \dingo & \dingoT & \dingoT \\
        & Baseline & Random & Data-based \\
        \midrule
        W/ glitch-affected events & 1.4\% & 2.5\% &  4.2\%\\
        W/o glitch-affected events & 1.8\% & 2.7\% & 5.2\% \\
        \bottomrule
    \end{tabularx}
\end{table}

\section{Validating \dingoT}
To ensure that the \dingoT model is well calibrated and aligns with standard methods, we provide P-P plots for each detector configuration and compare posteriors obtained with \dingoT to the standard sampling method Bilby~\citep{bilby}, commonly used in official LVK~analyses.

\sec{P-P plot per detector configuration}
In a P-P plot, the percentile rank of the true value within its posterior marginal is computed, and the cumulative distribution function~(CDF) of these ranks is visualized for each parameter.
A well-calibrated model yields uniformly distributed percentiles, producing a diagonal CDF.
Based on the posterior samples obtained for $10^3$~injections, we construct separate P–P~plots per detector configuration and include them in Fig.~\ref{fig:pp_plot_data}.
We also report the distribution of $p$-values from Kolmogorov–Smirnov tests to quantify deviations from uniformity and highlight the parameters where $p < 0.05$, indicating deviations larger than expected under the uniform assumption.
Overall, the \dingoT models demonstrate good calibration across all detector configurations on simulated data.

\begin{figure}[ht]
    \includegraphics[width=\columnwidth]{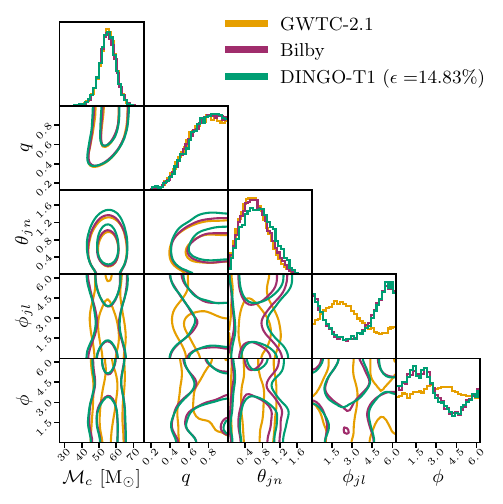}
    \caption{
    Comparison of \dingoT posteriors to GWTC-2.1 and Bilby results for the event GW190701\_203306.
    Only the parameters displaying the largest deviations between Bilby and \dingoT are included. The GWTC-2.1 posterior displays deviations due to different analysis settings such as the type of PSD, the reference frequency, and mixed posterior samples from different waveform models.
    }
    \label{fig:bilby_posterior_comparison}
\end{figure}

\sec{Comparison to Bilby and LVK catalogs}While we would ideally compare the \dingoT posteriors to the official LVK~catalog results, we do not expect perfect agreement between these results due to several reasons: 
(1)~The posterior samples provided by the~LVK are obtained using BayesWave PSDs~\citep{cornish:2014, littenberg:2014, cornish:2020}, while \dingoT is trained with Welch PSDs. Furthermore, we include recent changes regarding the window factor which can lead to biases in the posterior~\citep{talbot_window_fix:2025}.
(2)~Posterior samples from the catalogs are mixed results obtained from separate nested sampling runs using different waveform models and reference frequencies. 
Since we train \dingoT only on IMRPhenomXPHM, it cannot be guaranteed to obtain aligned posteriors. 
(3)~The prior ranges chosen for \dingoT might not be perfectly consistent with the prior ranges chosen for the analysis of a specific event in the LVK~catalogs. 
To still allow for a valid comparison, we obtain posterior samples with Bilby using the PSDs, waveform model, window factor correction, and prior ranges selected for \dingoT.
Since it would require significant compute resources to evaluate Bilby on all events reported in this \emph{Letter}, we select GW190701\_203306 as an example. 
In Fig.~\ref{fig:bilby_posterior_comparison}, we show a corner plot of the posterior parameters exhibiting the largest deviations between Bilby and \dingoT.
To illustrate that a direct comparison to GWTC-2.1 is not possible, we also include the official posterior samples from the LVK~data release~\citep{ligo_data_release_gwtc21_zenodo:2022}.

\section{Applications}
To demonstrate how the flexibility of \dingoT enables new studies and investigations, we select two examples applications:

\sec{Posteriors under different detectors}Since we can flexibly adapt the \dingoT model to a different set of interferometers at inference time, we analyze the 48~events with all possible detector combinations and show the overview in Tab.~\ref{table:efficiencies_all_det_configs}. 
The posterior change of an exemplary event is shown in Fig.~3a of the main text.
Such a study not only shows the consistent performance of \dingoT, but it can also provide additional insight into whether certain detectors are responsible for low sample efficiency.
For example, GW200129\_065458 exhibits low efficiency across all detector configurations involving the Livingston detector---consistent with the glitch subtraction applied in the official LVK~analysis.
Additionally, low sample efficiency in all detector configurations can point to mistakes in data pre-processing.
In an initial analysis of the event GW190517\_055101, the Hanford~PSD used for conditioning the \dingo models did not capture access power around~\SI{500}{\hertz}, resulting in $\epsilon = 0\%$ across all models and detector configurations involving this detector.
This failure case is resolved by either masking the affected frequency range or employing a more suitable PSD. The latter leads to an efficiency increase of up to 2.17\%.

\sec{Inspiral–merger–ringdown consistency tests}General relativity~(GR) can be tested by comparing the final mass and spin of the remnant black hole inferred from the low- and high-frequency parts of the signals~\citep{ghosh:2018, ligo_test_gr_gwtc2:2020, lvk_gwtc3_tests_gr:2025}.
For such inspiral–merger–ringdown~(IMR) consistency tests, the frequency-domain signal is divided into two parts defined by the cutoff frequency~$f_\mathrm{cut}$ and both segments---the inspiral part roughly corresponding to $[f_\mathrm{min}, f_\mathrm{cut}]$ and the postinspiral part corresponding to $[f_\mathrm{cut}, f_\mathrm{max}]$---are analyzed separately.
The cutoff frequency is defined as the dominant mode GW~frequency of a remnant Kerr black hole's innermost stable circular orbit~\citep{lvk_gwtc3_tests_gr:2025}.
To constrain deviations between the inspiral~(insp) and postinspiral~(postinsp), the following fractional deviations are defined for the final mass~$M_f$ and dimensionless spin~$\chi_f$:
\begin{equation}
	\frac{\Delta M_\text{f}}{\overline{M}_\text{f}} = 2\frac{M_\text{f}^\text{insp} -M_\text{f}^\text{postinsp} }{M_\text{f}^\text{insp} + M_\text{f}^\text{postinsp}}
\end{equation}
and 
\begin{equation}
	\frac{\Delta \chi_\text{f}}{\overline{\chi}_\text{f}} = 2\frac{\chi_\text{f}^\text{insp} -\chi_\text{f}^\text{postinsp} }{\chi_\text{f}^\text{insp} + \chi_\text{f}^\text{postinsp}}~,
\end{equation}
where $\overline{M}_\text{f}$ and $\overline{\chi}_\text{f}$ correspond to 
the mean of the final mass and spin analyzed with inspiral and postinspiral data.
The two dimensional posterior distribution for these fractional deviations is expected to peak at $(0,0)$ when the test is applied to a signal from a quasi-circular binary black hole coalescence in GR~\citep{ghosh:2018, ligo_test_gr_gwtc2:2020, lvk_gwtc3_tests_gr:2025}.

\definecolor{GW190408}{HTML}{3b4cc0}
\definecolor{GW190521}{HTML}{c0282f}
\definecolor{GW190630}{HTML}{afcafc}
\definecolor{GW190828}{HTML}{f6bea4}
\definecolor{GW200208}{HTML}{cb3e38}
\definecolor{GW200224}{HTML}{b40426}
\definecolor{GW200311}{HTML}{dcdddd}
\begin{table}[ht]
	\caption{
		Frequency settings for the inspiral-merger-ringdown consistency tests. 
        We provide the colors as a legend for Fig.~3b of the main text.
        }
	\label{table:IMR_consistency_tests}
	\centering
	\setlength{\tabcolsep}{6pt}
	\begin{tabularx}{0.93\columnwidth}{l c c c}
		\toprule
		\multirow{2}{*}{Event} & \multirow{2}{*}{Color} & \multicolumn{2}{c}{Frequency ranges [\si{\hertz}]}
		 \\
		 & & Inspiral & Postinspiral \\
		\midrule
        GW190408\_181802 & \textcolor{GW190408}{\rule[0.5ex]{1em}{1pt}} & [20,~164] & [164,~896]\\
        GW190521\_074359 & \textcolor{GW190521}{\rule[0.5ex]{1em}{1pt}} & [20,~105] & [105,~224]\\
        GW190630\_185205 & \textcolor{GW190630}{\rule[0.5ex]{1em}{1pt}} & [20,~135] & [135,~896]\\
		GW190828\_063205 & \textcolor{GW190828}{\rule[0.5ex]{1em}{1pt}} & [20,~132] & [132,~896]\\
		GW200208\_130117 & \textcolor{GW200208}{\rule[0.5ex]{1em}{1pt}} & [20,~~98] & [~98,~448]\\
		GW200224\_222234 & \textcolor{GW200224}{\rule[0.5ex]{1em}{1pt}} & [20,~107] & [107,~448]\\
        GW200311\_115853 & \textcolor{GW200311}{\rule[0.5ex]{1em}{1pt}} & [20,~122] & [102,~896] \\
		\bottomrule
	\end{tabularx}
\end{table}

To demonstrate that we can employ the flexibility of the \dingoT model for in-depth studies, we investigate two separate objectives:

We first perform IMR consistency tests for seven events listed in Tab.~\ref{table:IMR_consistency_tests}. 
These were selected from the 12~events analyzed in~\citep{lvk_gwtc3_tests_gr:2025} based on compatibility with the training priors and the requirement that more than 5,000~effective samples could be obtained with \dingoT.
We separately analyze the inspiral and postinspiral signals with the \dingoT model and draw 5,000~effective samples for both posteriors.
The resulting 2D~posterior distributions of fractional deviations are shown in Fig.~3b in the main text which are not reweighed to a uniform prior as in~\citep{lvk_gwtc3_tests_gr:2025} and thus not directly comparable to the results shown in~\citep{lvk_gwtc3_tests_gr:2025}.

\begin{figure}[ht]
    \includegraphics[width=\columnwidth]{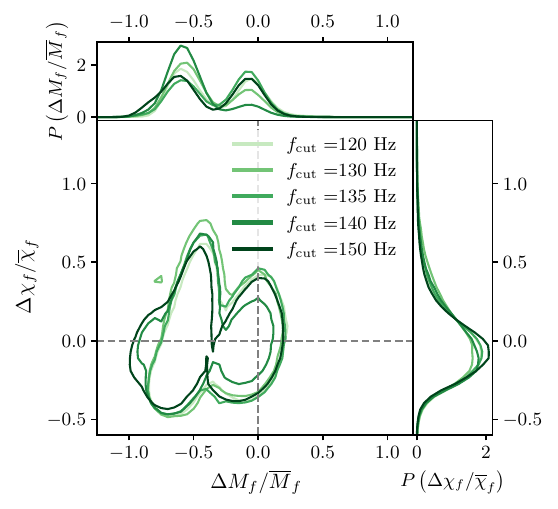}
    \caption{
    Inspiral-merger-ringdown consistency test for GW190630\_185205 where the signal was analyzed with the \dingoT model on the inspiral and postinspiral part of the waveform separated by different $f_\mathrm{cut}$ values. The main panel shows the 90\% credible regions of the 2D posteriors on $(\Delta M_f / \overline{M}_f, \Delta \chi_f / \overline{\chi}_f)$. The side panels show the marginalized posterior of $\Delta M_f / \overline{M}_f$ and $\Delta \chi_f / \overline{\chi}_f$.}
    \label{fig:imrc_test_f_cut}
\end{figure}

\begin{figure*}[ht]
    \centering
    \includegraphics[width=\textwidth]{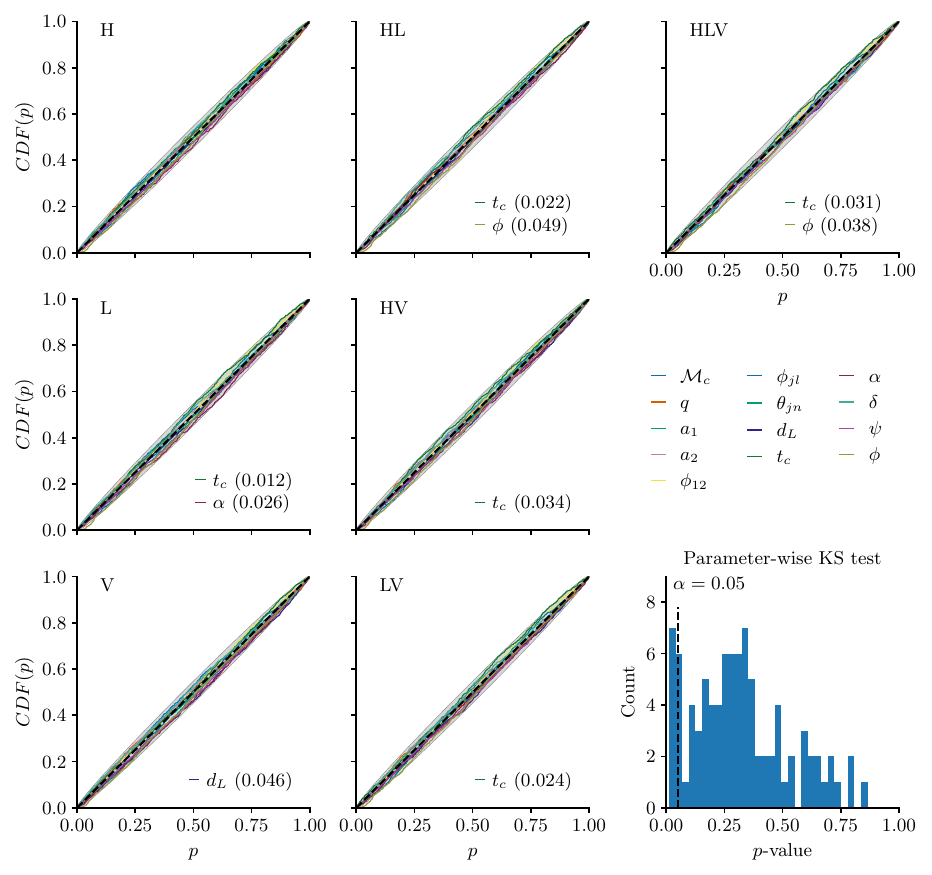}
    \caption{
    P-P plot for 1000 injections evaluated with different detector configurations for \dingoT. The parameters where the Kolmogorov-Smirnov~(KS) test falls below $\alpha = 0.05$ are included explicitly and the overall distribution of KS-based $p$-values is shown in a histogram. We show the $3\sigma$ confidence intervals as a gray shaded band.
    }
    \label{fig:pp_plot_data}
\end{figure*}

Second, we repeat the IMR~consistency test for GW190630\_185205 with different frequency cuts to investigate the robustness of our previous result to varying assumptions about~$f_\mathrm{cut}$~\citep{ghosh:2018}.
In addition to $f_\mathrm{cut} = \SI{135}{\hertz}$, we choose $120, 130, 140,$ and $\SI{150}{\hertz}$ and show the results in Fig.~\ref{fig:imrc_test_f_cut}.
While we observe small deviations in the posterior distribution, the result for $f_\mathrm{cut} = \SI{135}{\hertz}$ is consistent with the results for different frequency cuts.

This flexibility has the potential to reduce the computational cost of IMR~consistency tests and facilitate detailed robustness studies.

\section{Computational advantage of \dingoT}
By training a single flexible \dingoT model, we avoid training 94~separate \dingo models. 
This large number of models would be necessary since frequency ranges and detector combinations vary between experiments.
Specifically, seven \dingo models would be required to perform the injection study with different detector combinations on the full frequency domain (Fig.~\ref{fig:injections_violin_plot}), 65~models would have to be trained to cover all detector combinations and frequency settings for each event (Tab.~\ref{table:efficiencies_all_det_configs}), two models are needed for each of the seven IMR~consistency tests (Tab.~\ref{table:IMR_consistency_tests}), and two specific models would have to be trained for each IMR consistency test where we investigate four different $f_\mathrm{cut}$ values (Fig.~\ref{fig:imrc_test_f_cut}). 
This results in a total of $7 + 65 + 2~\cdot~7 + 2~\cdot~4 = 94$ baseline \dingo models which we can replace by a single \dingoT model.

\FloatBarrier

\begin{table*}[ht]
    \caption{Detector and frequency configurations for 48~O3~events. Events with changes to~$f_\mathrm{min}$ are highlighted in bold.
    Furthermore, we highlight events where different glitch mitigation techniques were applied to a specific detector: 
    $*$~glitch subtraction, 
    $\dagger$~update of $f_\mathrm{min}$, $\diamond$~BayesWave deglitching, 
    and $\ddagger$~linear subtraction. Details are provided in Tab.~\ref{table:glitch_mitigation}.
    }
    \label{table:event_configs}
    \centering
    \setlength{\tabcolsep}{6pt}
    \begin{tabularx}{0.84\linewidth}{l c ccc r cc l}
        \toprule
        \multirow{2}{*}{Event} & 
        \multirow{2}{*}{Detectors} & 
        \multicolumn{3}{c}{$f_\mathrm{min}$ [Hz]} & 
        \multirow{2}{*}{\shortstack{$f_\mathrm{max}$[Hz] \\ {H/L/V}}} & 
        \multicolumn{2}{c}{PSD Notching (V)} & 
        \multirow{2}{*}{Catalog} \\
        & & H & L & V & & $f_\mathrm{min}$[Hz] & $f_\mathrm{max}$[Hz] & \\
        \midrule
        \begin{filecontents*}{table_event_configs.csv}
Event,ifos,minimum-frequency H1,minimum-frequency L1,minimum-frequency V1,maximum-frequency,psd-notching minimum-frequency V1,psd-notching maximum-frequency V1,catalog
\text{GW190408\_181802},HLV,20,20,20,896,-,-,GWTC-2.1
\text{GW190413\_052954},HLV,20,20,20,896,-,-,GWTC-2.1
\textbf{GW190413\_134308}\text{$^{*\dagger \mathrm{L}}$},HLV,20,$\mathbf{35}$,20,448,-,-,GWTC-2.1
\text{GW190421\_213856},HL,20,20,-,448,-,-,GWTC-2.1
\text{GW190426\_190642},HLV,20,20,20,448,-,-,GWTC-2.1
\text{GW190503\_185404$^{* \mathrm{L}}$},HLV,20,20,20,896,-,-,GWTC-2.1
\text{GW190513\_205428$^{* \mathrm{L}}$},HLV,20,20,20,896,-,-,GWTC-2.1
\textbf{GW190514\_065416}\text{$^{*\dagger \mathrm{L}}$},HL,20,$\mathbf{50}$,-,448,-,-,GWTC-2.1
\text{GW190517\_055101},HLV,20,20,20,896,-,-,GWTC-2.1
\text{GW190519\_153544},HLV,20,20,20,448,-,-,GWTC-2.1
\text{GW190521\_074359},HL,20,20,-,224,-,-,GWTC-2.1
\text{GW190527\_092055},HL,20,20,-,896,-,-,GWTC-2.1
\text{GW190602\_175927},HLV,20,20,20,448,-,-,GWTC-2.1
\text{GW190620\_030421},LV,-,20,20,448,-,-,GWTC-2.1
\text{GW190630\_185205},LV,-,20,20,896,-,-,GWTC-2.1
\text{GW190701\_203306$^{* \mathrm{L}}$},HLV,20,20,20,448,-,-,GWTC-2.1
\text{GW190706\_222641},HLV,20,20,20,896,-,-,GWTC-2.1
\text{GW190719\_215514},HL,20,20,-,896,-,-,GWTC-2.1
\textbf{GW190727\_060333}\text{$^{\dagger \mathrm{L}}$},HLV,20,$\mathbf{50}$,20,896,-,-,GWTC-2.1
\text{GW190731\_140936},HL,20,20,-,896,-,-,GWTC-2.1
\text{GW190803\_022701},HLV,20,20,20,896,-,-,GWTC-2.1
\text{GW190828\_063405},HLV,20,20,20,896,-,-,GWTC-2.1
\text{GW190910\_112807},LV,-,20,20,448,-,-,GWTC-2.1
\text{GW190915\_235702},HLV,20,20,20,896,-,-,GWTC-2.1
\text{GW190916\_200658},HLV,20,20,20,896,-,-,GWTC-2.1
\text{GW190925\_232845},HV,20,-,20,1792,-,-,GWTC-2.1
\text{GW190926\_050336},HLV,20,20,20,896,-,-,GWTC-2.1
\text{GW190929\_012149},HLV,20,20,20,896,-,-,GWTC-2.1
\text{GW191109\_010717$^{\diamond \mathrm{HL}}$},HL,20,20,-,448,-,-,GWTC-3
\text{GW191127\_050227$^{\diamond \mathrm{H}}$},HLV,20,20,20,896,46,51,GWTC-3
\text{GW191204\_110529},HL,20,20,-,896,-,-,GWTC-3
\text{GW191215\_223052},HLV,20,20,20,896,46,51,GWTC-3
\text{GW191222\_033537},HL,20,20,-,448,-,-,GWTC-3
\text{GW191230\_180458},HLV,20,20,20,896,46,51,GWTC-3
\text{GW200112\_155838},LV,-,20,20,896,46,51,GWTC-3
\text{GW200128\_022011},HL,20,20,-,448,-,-,GWTC-3
\text{GW200129\_065458$^{\ddagger \mathrm{L}}$},HLV,20,20,20,896,46,51,GWTC-3
\text{GW200208\_130117},HLV,20,20,20,448,46,51,GWTC-3
\text{GW200208\_222617},HLV,20,20,20,1792,46,51,GWTC-3
\text{GW200209\_085452},HLV,20,20,20,896,46,51,GWTC-3
\text{GW200216\_220804},HLV,20,20,20,896,46,51,GWTC-3
\text{GW200219\_094415},HLV,20,20,20,896,46,51,GWTC-3
\text{GW200220\_061928},HLV,20,20,20,224,46,51,GWTC-3
\text{GW200220\_124850},HL,20,20,-,896,-,-,GWTC-3
\text{GW200224\_222234},HLV,20,20,20,448,46,51,GWTC-3
\text{GW200302\_015811},HV,20,-,20,896,46,51,GWTC-3
\text{GW200306\_093714},HL,20,20,-,896,-,-,GWTC-3
\text{GW200311\_115853},HLV,20,20,20,896,46,51,GWTC-3
\end{filecontents*}
\csvreader[late after line=\\,late after last line=\\\bottomrule]{table_event_configs.csv}
          {}
          {\csvcoli & \csvcolii & \csvcoliii & \csvcoliv & \csvcolv & \csvcolvi & \csvcolvii & \csvcolviii & \csvcolix}
    \end{tabularx}
\end{table*}

\begin{table*}[ht]
    \caption{
        Overview of events from O3~\citep{ligo_gwtc-21:2024, ligo_gwtc-3:2023} where different glitch mitigation techniques are applied.
    }
    \label{table:glitch_mitigation}
    \centering
    \setlength{\tabcolsep}{6pt}
    \begin{tabularx}{0.97\textwidth}{lccccl}
        \toprule
        \multirow{2}{*}{Event} & \multirow{2}{*}{Detectors} & \multirow{2}{*}{Mitigation} & Affected & Frequency & \multirow{2}{*}{Catalog} \\ 
         & & & detectors & update & \\ \midrule GW190413\_134308$^{*\dagger\mathrm{L}}$ & HLV & glitch subtraction, glitch-only model & L & $f_\mathrm{min} = \SI{35}{\hertz}$ & GWTC-2.1 \\
         GW190503\_185404$^{*\mathrm{L}}$ & HLV & glitch subtraction, glitch-only model & L & - & GWTC-2.1 \\
        GW190513\_205428$^{*\mathrm{L}}$ & HLV & glitch subtraction, glitch-only model & L & - & GWTC-2.1 \\
        GW190514\_065416$^{*\dagger\mathrm{L}}$ & HL & glitch subtraction, glitch-only model & L & $f_\mathrm{min} = \SI{50}{\hertz}$ & GWTC-2.1 \\
        GW190701\_203306$^{*\mathrm{L}}$ & HLV & glitch subtraction, glitch+signal model & L & - & GWTC-2.1 \\
        GW190727\_060333$^{\dagger\mathrm{L}}$ & HLV & - & L & $f_\mathrm{min} = \SI{50}{\hertz}$ & GWTC-2.1 \\
        GW191109\_010717$^{\diamond\mathrm{HL}}$ & HL & BayesWave deglitching & HL & - & GWTC-3\\
        GW191127\_050227$^{\diamond\mathrm{H}}$ & HLV & BayesWave deglitching & H & - & GWTC-3 \\
        GW200129\_065458$^{\ddagger\mathrm{L}}$ & HLV & Linear subtraction & L & - & GWTC-3 \\
        \bottomrule
    \end{tabularx}
\end{table*}

\begin{table*}[ht]
    \centering
    \caption{
        Sample efficiencies for 48 O3~events with 17~different data analysis settings evaluated with \dingoT.
        The \dingo baseline is evaluated on HLV events without adapting the frequency domain.
        We highlight events where different glitch mitigation techniques were applied to a specific detector: 
        $*$~glitch subtraction, 
        $\dagger$~update of $f_\mathrm{min}$, $\diamond$~BayesWave deglitching, 
        and $\ddagger$~linear subtraction. Details are provided in Tab.~\ref{table:glitch_mitigation}.
    }
    \label{table:efficiencies_events}
    \begin{minipage}{0.49\textwidth}
        \centering
        \small
        \setlength{\tabcolsep}{0pt}
        \begin{tabularx}{\textwidth}{lcccc}
            \toprule
            \multirow{2}{*}{Event} & 
            \multirow{2}{*}{Det.~~~} & 
            \multirow{2}{*}{\shortstack{\dingo \\ Baseline}} & 
            \multicolumn{2}{c}{\dingoT} \\
            & & & Random & Data-based \\ 
            \midrule
            \begin{filecontents*}{table_efficiencies_1.csv}
Event,Detectors,baseline,random,data
\text{GW190408\_181802},HLV,\text{\gradient{0.63}},\text{\gradient{0.73}},\text{\gradient{8.07}}
\text{GW190413\_052954},HLV,\text{\gradient{5.38}},\text{\gradient{9.33}},\text{\gradient{11.77}}
\text{GW190413\_134308}$^{*\dagger \mathrm{L}}$,HLV,\text{\gradient{1.19}},\text{\gradient{5.2}},\text{\gradient{4.52}}
\text{GW190426\_190642},HLV,\text{\gradient{0.97}},\text{\gradient{0.39}},\text{\gradient{0.79}}
\text{GW190503\_185404}$^{* \mathrm{L}}$,HLV,\text{\gradient{0.18}},\text{\gradient{0.5}},\text{\gradient{1.57}}
\text{GW190513\_205428}$^{* \mathrm{L}}$,HLV,\text{\gradient{0.2}},\text{\gradient{1.54}},\text{\gradient{0.16}}
\text{GW190517\_055101},HLV,\text{\gradient{0.06}},\text{\gradient{1.61}},\text{\gradient{2.17}}
\text{GW190519\_153544},HLV,\text{\gradient{1.87}},\text{\gradient{1.93}},\text{\gradient{4.26}}
\text{GW190602\_175927},HLV,\text{\gradient{10.33}},\text{\gradient{13.13}},\text{\gradient{13.09}}
\text{GW190701\_203306}$^{* \mathrm{L}}$,HLV,\text{\gradient{2.17}},\text{\gradient{2.81}},\text{\gradient{14.83}}
\text{GW190706\_222641},HLV,\text{\gradient{3.5}},\text{\gradient{2.35}},\text{\gradient{2.49}}
\text{GW190727\_060333}$^{\dagger \mathrm{L}}$,HLV,\text{\gradient{0.77}},\text{\gradient{1.41}},\text{\gradient{0.84}}
\text{GW190803\_022701},HLV,\text{\gradient{11.55}},\text{\gradient{16.86}},\text{\gradient{16.39}}
\text{GW190828\_063405},HLV,\text{\gradient{0.92}},\text{\gradient{4.27}},\text{\gradient{3.99}}
\text{GW190915\_235702},HLV,\text{\gradient{3.42}},\text{\gradient{5.29}},\text{\gradient{8.23}}
\text{GW190916\_200658},HLV,\text{\gradient{11.0}},\text{\gradient{19.37}},\text{\gradient{19.41}}
\text{GW190926\_050336},HLV,\text{\gradient{1.23}},\text{\gradient{3.52}},\text{\gradient{1.63}}
\text{GW190929\_012149},HLV,\text{\gradient{1.64}},\text{\gradient{1.16}},\text{\gradient{3.04}}
\text{GW191127\_050227}$^{\diamond \mathrm{H}}$,HLV,\text{\gradient{0.16}},\text{\gradient{0.4}},\text{\gradient{0.28}}
\text{GW191215\_223052},HLV,\text{\gradient{0.07}},\text{\gradient{0.2}},\text{\gradient{4.05}}
\text{GW191230\_180458},HLV,\text{\gradient{5.5}},\text{\gradient{9.83}},\text{\gradient{14.61}}
\text{GW200129\_065458}$^{\ddagger \mathrm{L}}$,HLV,\text{\gradient{0.05}},\text{\gradient{0.01}},\text{\gradient{0.29}}
\text{GW200208\_130117},HLV,\text{\gradient{1.46}},\text{\gradient{2.12}},\text{\gradient{11.58}}
\text{GW200208\_222617},HLV,\text{\gradient{2.49}},\text{\gradient{2.71}},\text{\gradient{2.86}}
\end{filecontents*}
\csvreader[late after line=\\,late after last line=\\\bottomrule]{table_efficiencies_1.csv}
              {}
              {\csvcoli & \csvcolii & \csvcoliii & \csvcoliv & \csvcolv}
        \end{tabularx}
    \end{minipage}
    \hfill
    \begin{minipage}{0.49\textwidth}
        \centering
        \small
        \setlength{\tabcolsep}{0pt}
        \begin{tabularx}{\textwidth}{lcccc}
            \toprule
            \multirow{2}{*}{Event} & 
            \multirow{2}{*}{Det.~~~} & 
            \multirow{2}{*}{\shortstack{\dingo \\ Baseline}} & 
            \multicolumn{2}{c}{\dingoT} \\
            & & & Random & Data-based \\ 
            \midrule
            \begin{filecontents*}{table_efficiencies_2.csv}
Event,Detectors,baseline,random,data
\text{GW200209\_085452},HLV,\text{\gradient{1.39}},\text{\gradient{3.97}},\text{\gradient{2.52}}
\text{GW200216\_220804},HLV,\text{\gradient{5.31}},\text{\gradient{4.67}},\text{\gradient{2.2}}
\text{GW200219\_094415},HLV,\text{\gradient{4.09}},\text{\gradient{3.11}},\text{\gradient{5.12}}
\text{GW200220\_061928},HLV,\text{\gradient{0.62}},\text{\gradient{1.81}},\text{\gradient{10.17}}
\text{GW200224\_222234},HLV,\text{\gradient{0.23}},\text{\gradient{2.67}},\text{\gradient{5.49}}
\text{GW200311\_115853},HLV,\text{\gradient{1.83}},\text{\gradient{6.59}},\text{\gradient{9.2}}
\text{GW190421\_213856},HL,-,\text{\gradient{9.33}},\text{\gradient{17.8}}
\text{GW190514\_065416}$^{*\dagger \mathrm{L}}$,HL,-,\text{\gradient{6.83}},\text{\gradient{15.22}}
\text{GW190521\_074359},HL,-,\text{\gradient{0.17}},\text{\gradient{1.28}}
\text{GW190527\_092055},HL,-,\text{\gradient{7.2}},\text{\gradient{9.04}}
\text{GW190719\_215514},HL,-,\text{\gradient{6.26}},\text{\gradient{14.56}}
\text{GW190731\_140936},HL,-,\text{\gradient{26.71}},\text{\gradient{18.22}}
\text{GW191109\_010717}$^{\diamond \mathrm{HL}}$,HL,-,\text{\gradient{0.19}},\text{\gradient{1.52}}
\text{GW191204\_110529},HL,-,\text{\gradient{0.11}},\text{\gradient{0.07}}
\text{GW191222\_033537},HL,-,\text{\gradient{16.03}},\text{\gradient{15.61}}
\text{GW200128\_022011},HL,-,\text{\gradient{4.49}},\text{\gradient{8.28}}
\text{GW200220\_124850},HL,-,\text{\gradient{3.83}},\text{\gradient{12.94}}
\text{GW200306\_093714},HL,-,\text{\gradient{0.57}},\text{\gradient{0.09}}
\text{GW190925\_232845},HV,-,\text{\gradient{0.06}},\text{\gradient{0.86}}
\text{GW200302\_015811},HV,-,\text{\gradient{0.5}},\text{\gradient{1.4}}
\text{GW190620\_030421},LV,-,\text{\gradient{0.63}},\text{\gradient{0.81}}
\text{GW190630\_185205},LV,-,\text{\gradient{0.61}},\text{\gradient{1.15}}
\text{GW190910\_112807},LV,-,\text{\gradient{0.73}},\text{\gradient{3.22}}
\text{GW200112\_155838},LV,-,\text{\gradient{1.67}},\text{\gradient{5.79}}
\end{filecontents*}
\csvreader[late after line=\\,late after last line=\\\bottomrule]{table_efficiencies_2.csv}
              {}
              {\csvcoli & \csvcolii & \csvcoliii & \csvcoliv & \csvcolv}
        \end{tabularx}
    \end{minipage}
\end{table*}

\begin{table*}[ht]
    \caption{
        Sample efficiencies for events evaluated on all potential detector configurations with \dingoT. We highlight events where different glitch mitigation techniques were applied to a specific detector: 
        $*$~glitch subtraction, 
        $\dagger$~update of $f_\mathrm{min}$, $\diamond$~BayesWave deglitching, 
        and $\ddagger$~linear subtraction. Details are provided in Table~\ref{table:glitch_mitigation}.
    }
    \label{table:efficiencies_all_det_configs}
    \centering
    \begin{tabularx}{0.86\linewidth}{lcc@{}c@{}c@{}c@{}c@{}c@{}c@{}}
        \toprule
        Event & Detectors & HLV & HL & HV & LV & H & L & V \\ \midrule
        \begin{filecontents*}{table_all_detector_configs_efficiencies_04_data_based_larger_fmasking.csv}
Event,Detectors,HLV,HL,HV,LV,H,L,V
\text{GW190408\_181802},HLV,\text{\gradient{8.07}},\text{\gradient{8.28}},\text{\gradient{0.64}},\text{\gradient{1.35}},\text{\gradient{2.79}},\text{\gradient{6.71}},\text{\gradient{62.23}}
\text{GW190413\_052954},HLV,\text{\gradient{11.77}},\text{\gradient{9.52}},\text{\gradient{2.41}},\text{\gradient{0.74}},\text{\gradient{9.5}},\text{\gradient{18.61}},\text{\gradient{37.48}}
\text{GW190413\_134308}$^{*\dagger \mathrm{L}}$,HLV,\text{\gradient{4.52}},\text{\gradient{7.84}},\text{\gradient{8.29}},\text{\gradient{1.36}},\text{\gradient{21.16}},\text{\gradient{11.78}},\text{\gradient{32.2}}
\text{GW190426\_190642},HLV,\text{\gradient{0.79}},\text{\gradient{2.27}},\text{\gradient{7.41}},\text{\gradient{0.0}},\text{\gradient{29.53}},\text{\gradient{0.04}},\text{\gradient{49.52}}
\text{GW190503\_185404}$^{* \mathrm{L}}$,HLV,\text{\gradient{1.57}},\text{\gradient{1.31}},\text{\gradient{1.0}},\text{\gradient{0.41}},\text{\gradient{17.09}},\text{\gradient{2.14}},\text{\gradient{60.85}}
\text{GW190513\_205428}$^{* \mathrm{L}}$,HLV,\text{\gradient{0.16}},\text{\gradient{0.52}},\text{\gradient{4.19}},\text{\gradient{2.04}},\text{\gradient{11.74}},\text{\gradient{4.17}},\text{\gradient{22.9}}
\text{GW190517\_055101},HLV,\text{\gradient{2.17}},\text{\gradient{4.28}},\text{\gradient{1.79}},\text{\gradient{0.27}},\text{\gradient{1.36}},\text{\gradient{0.36}},\text{\gradient{13.06}}
\text{GW190519\_153544},HLV,\text{\gradient{4.26}},\text{\gradient{0.85}},\text{\gradient{3.71}},\text{\gradient{4.88}},\text{\gradient{3.38}},\text{\gradient{9.09}},\text{\gradient{59.66}}
\text{GW190602\_175927},HLV,\text{\gradient{13.09}},\text{\gradient{13.18}},\text{\gradient{28.09}},\text{\gradient{5.4}},\text{\gradient{27.24}},\text{\gradient{15.95}},\text{\gradient{26.58}}
\text{GW190701\_203306}$^{* \mathrm{L}}$,HLV,\text{\gradient{14.83}},\text{\gradient{14.05}},\text{\gradient{9.13}},\text{\gradient{5.45}},\text{\gradient{31.43}},\text{\gradient{5.93}},\text{\gradient{38.78}}
\text{GW190706\_222641},HLV,\text{\gradient{2.49}},\text{\gradient{5.65}},\text{\gradient{13.73}},\text{\gradient{6.21}},\text{\gradient{0.1}},\text{\gradient{10.14}},\text{\gradient{27.66}}
\text{GW190727\_060333}$^{\dagger \mathrm{L}}$,HLV,\text{\gradient{0.84}},\text{\gradient{17.6}},\text{\gradient{2.95}},\text{\gradient{3.29}},\text{\gradient{5.09}},\text{\gradient{0.45}},\text{\gradient{23.71}}
\text{GW190803\_022701},HLV,\text{\gradient{16.39}},\text{\gradient{28.23}},\text{\gradient{19.48}},\text{\gradient{9.04}},\text{\gradient{28.74}},\text{\gradient{7.89}},\text{\gradient{42.92}}
\text{GW190828\_063405},HLV,\text{\gradient{3.99}},\text{\gradient{17.35}},\text{\gradient{8.7}},\text{\gradient{7.11}},\text{\gradient{24.58}},\text{\gradient{9.26}},\text{\gradient{64.41}}
\text{GW190915\_235702},HLV,\text{\gradient{8.23}},\text{\gradient{13.56}},\text{\gradient{0.89}},\text{\gradient{5.24}},\text{\gradient{1.9}},\text{\gradient{14.02}},\text{\gradient{42.52}}
\text{GW190916\_200658},HLV,\text{\gradient{19.41}},\text{\gradient{20.19}},\text{\gradient{5.16}},\text{\gradient{18.13}},\text{\gradient{11.2}},\text{\gradient{27.25}},\text{\gradient{42.29}}
\text{GW190926\_050336},HLV,\text{\gradient{1.63}},\text{\gradient{4.25}},\text{\gradient{11.17}},\text{\gradient{12.38}},\text{\gradient{16.39}},\text{\gradient{19.97}},\text{\gradient{53.75}}
\text{GW190929\_012149},HLV,\text{\gradient{3.04}},\text{\gradient{4.74}},\text{\gradient{10.61}},\text{\gradient{0.0}},\text{\gradient{24.68}},\text{\gradient{0.01}},\text{\gradient{50.24}}
\text{GW191127\_050227}$^{\diamond \mathrm{H}}$,HLV,\text{\gradient{0.28}},\text{\gradient{0.97}},\text{\gradient{8.21}},\text{\gradient{5.15}},\text{\gradient{16.13}},\text{\gradient{28.83}},\text{\gradient{6.04}}
\text{GW191215\_223052},HLV,\text{\gradient{4.05}},\text{\gradient{5.79}},\text{\gradient{5.13}},\text{\gradient{1.25}},\text{\gradient{6.55}},\text{\gradient{2.02}},\text{\gradient{56.09}}
\text{GW191230\_180458},HLV,\text{\gradient{14.61}},\text{\gradient{8.4}},\text{\gradient{2.68}},\text{\gradient{16.51}},\text{\gradient{7.79}},\text{\gradient{16.88}},\text{\gradient{58.75}}
\text{GW200129\_065458}$^{\ddagger \mathrm{L}}$,HLV,\text{\gradient{0.29}},\text{\gradient{0.29}},\text{\gradient{2.16}},\text{\gradient{0.01}},\text{\gradient{14.15}},\text{\gradient{0.07}},\text{\gradient{2.66}}
\text{GW200208\_130117},HLV,\text{\gradient{11.58}},\text{\gradient{12.15}},\text{\gradient{5.17}},\text{\gradient{5.99}},\text{\gradient{32.27}},\text{\gradient{8.9}},\text{\gradient{17.75}}
\text{GW200208\_222617},HLV,\text{\gradient{2.86}},\text{\gradient{2.47}},\text{\gradient{4.73}},\text{\gradient{0.05}},\text{\gradient{5.55}},\text{\gradient{18.82}},\text{\gradient{18.48}}
\text{GW200209\_085452},HLV,\text{\gradient{2.52}},\text{\gradient{11.06}},\text{\gradient{3.65}},\text{\gradient{16.62}},\text{\gradient{9.66}},\text{\gradient{34.08}},\text{\gradient{41.28}}
\text{GW200216\_220804},HLV,\text{\gradient{2.2}},\text{\gradient{12.56}},\text{\gradient{12.41}},\text{\gradient{7.44}},\text{\gradient{21.27}},\text{\gradient{10.96}},\text{\gradient{52.1}}
\text{GW200219\_094415},HLV,\text{\gradient{5.12}},\text{\gradient{7.36}},\text{\gradient{3.27}},\text{\gradient{2.77}},\text{\gradient{11.76}},\text{\gradient{11.09}},\text{\gradient{1.9}}
\text{GW200220\_061928},HLV,\text{\gradient{10.17}},\text{\gradient{18.1}},\text{\gradient{6.07}},\text{\gradient{7.68}},\text{\gradient{8.67}},\text{\gradient{11.95}},\text{\gradient{61.93}}
\text{GW200224\_222234},HLV,\text{\gradient{5.49}},\text{\gradient{7.18}},\text{\gradient{1.65}},\text{\gradient{6.85}},\text{\gradient{10.2}},\text{\gradient{20.89}},\text{\gradient{4.01}}
\text{GW200311\_115853},HLV,\text{\gradient{9.2}},\text{\gradient{5.56}},\text{\gradient{11.57}},\text{\gradient{4.81}},\text{\gradient{7.71}},\text{\gradient{4.23}},\text{\gradient{3.64}}
\text{GW190421\_213856},HL,-,\text{\gradient{17.8}},-,-,\text{\gradient{10.08}},\text{\gradient{17.4}},-
\text{GW190514\_065416}$^{*\dagger \mathrm{L}}$,HL,-,\text{\gradient{15.22}},-,-,\text{\gradient{19.48}},\text{\gradient{4.83}},-
\text{GW190521\_074359},HL,-,\text{\gradient{1.28}},-,-,\text{\gradient{7.24}},\text{\gradient{1.26}},-
\text{GW190527\_092055},HL,-,\text{\gradient{9.04}},-,-,\text{\gradient{2.95}},\text{\gradient{5.56}},-
\text{GW190719\_215514},HL,-,\text{\gradient{14.56}},-,-,\text{\gradient{17.56}},\text{\gradient{0.05}},-
\text{GW190731\_140936},HL,-,\text{\gradient{18.22}},-,-,\text{\gradient{38.6}},\text{\gradient{4.69}},-
\text{GW191109\_010717}$^{\diamond \mathrm{HL}}$,HL,-,\text{\gradient{1.52}},-,-,\text{\gradient{0.11}},\text{\gradient{5.15}},-
\text{GW191204\_110529},HL,-,\text{\gradient{0.07}},-,-,\text{\gradient{19.49}},\text{\gradient{13.28}},-
\text{GW191222\_033537},HL,-,\text{\gradient{15.61}},-,-,\text{\gradient{6.84}},\text{\gradient{16.18}},-
\text{GW200128\_022011},HL,-,\text{\gradient{8.28}},-,-,\text{\gradient{38.67}},\text{\gradient{18.79}},-
\text{GW200220\_124850},HL,-,\text{\gradient{12.94}},-,-,\text{\gradient{23.86}},\text{\gradient{15.56}},-
\text{GW200306\_093714},HL,-,\text{\gradient{0.09}},-,-,\text{\gradient{6.0}},\text{\gradient{0.08}},-
\text{GW190925\_232845},HV,-,-,\text{\gradient{0.86}},-,\text{\gradient{8.16}},-,\text{\gradient{16.97}}
\text{GW200302\_015811},HV,-,-,\text{\gradient{1.4}},-,\text{\gradient{2.25}},-,\text{\gradient{31.8}}
\text{GW190620\_030421},LV,-,-,-,\text{\gradient{0.81}},-,\text{\gradient{1.5}},\text{\gradient{55.53}}
\text{GW190630\_185205},LV,-,-,-,\text{\gradient{1.15}},-,\text{\gradient{7.98}},\text{\gradient{16.08}}
\text{GW190910\_112807},LV,-,-,-,\text{\gradient{3.22}},-,\text{\gradient{4.2}},\text{\gradient{19.1}}
\text{GW200112\_155838},LV,-,-,-,\text{\gradient{5.79}},-,\text{\gradient{16.7}},\text{\gradient{20.93}}
\end{filecontents*}
\csvreader[late after line=\\,late after last line=\\\bottomrule]{table_all_detector_configs_efficiencies_04_data_based_larger_fmasking.csv}
          {}
          {\csvcoli & \csvcolii & \csvcoliii & \csvcoliv & \csvcolv & \csvcolvi & \csvcolvii & \csvcolviii & \csvcolix}
    \end{tabularx}
\end{table*}

	\bibliography{mybib.bib}
	
\end{document}